\documentclass[twocolumn,prb,showpacs]{revtex4}

\usepackage{graphicx}
\usepackage{dcolumn}
\usepackage{float}
\usepackage{amsmath}

\newcommand{\comment}[1]{}

\begin{document}


\title{Effect of electron-phonon interaction on spectroscopies
in graphene}

\author{J.P. Carbotte}
\affiliation{Department of Physics and Astronomy, McMaster University,
Hamilton, Ontario, Canada L8S 4M1} 
\affiliation{The Canadian
Institute for Advanced Research, Toronto, Ontario, Canada M5G 1Z8}
\author{E.J. Nicol}
\affiliation{Department of Physics, University of Guelph,
Guelph, Ontario, Canada N1G 2W1}
\email{nicol@physics.uoguelph.ca}
\author{S.G. Sharapov}
\affiliation{Bogolyubov Institute for Theoretical Physics, 14-b, Metrologicheskaya St., Kiev, 03680 Ukraine}

\date{\today}

\begin{abstract}
{We calculate the effect of the electron-phonon interaction on the 
electronic density of states (DOS), the quasiparticle properties and on the
optical conductivity of graphene. 
In metals with DOS constant on the
scale of
phonon energies, the electron-phonon renormalizations drop out of the
dressed
DOS, however, due to the Dirac nature of the electron dynamics in
graphene,
the band DOS is linear in energy and phonon structures remain, 
which can be
emphasized by taking an energy derivative.
There is a shift  in the chemical
potential and in the
position in energy of the Dirac point. Also,
the DOS can be changed from a linear dependence out of value
zero at the Dirac point
to quadratic out of a finite value. 
The optical scattering rate $1/\tau$ sets the energy scale for the rise
of the optical conductivity from its universal DC value $4e^2/\pi h$
(expected in the simplest theory
 when chemical potential and temperature are both $\ll 1/2\tau$) to its
universal AC background value $(\sigma_0=\pi e^2/2h)$.
As in ordinary metals
the DC conductivity remains unrenormalized while its AC value is changed.
The optical spectral weight under the intraband Drude is reduced
by a mass renormalization factor 
as is the effective scattering rate. Optical weight is
transferred to an Holstein
phonon-assisted side band. Due to Pauli blocking
the interband transitions are sharply suppressed, but also nearly
constant, below twice
the value of renormalized chemical potential
and also exhibit a phonon-assisted contribution. The universal
background conductivity is reduced below $\sigma_0$ at large
energies.
 
}
\end{abstract}


\maketitle

\section{Introduction}

Graphene consists of a monolayer of a honeycomb lattice of  carbon atoms.
It has been studied theoretically since the early work of Wallace.\cite{Wallace}
An important aspect of the charge dynamics in graphene is that it is governed by
a Dirac rather than a Schr\"odinger equation. Fermionic energies are proportional
to momentum with the effective speed of light $v_0\simeq
10^6$m/s. While 
the material was isolated
experimentally only in 2004,\cite{geimscience,novoselov}
many fascinating discoveries have already appeared\cite{geimnature,kimnature,geimnatmat}
including an integer quantum Hall effect with half integer filling factors,
a minimum conductivity and a Berry phase shift of $\pi$.\cite{geimnatmat,netormp,cormier,gusyninrev}

An important dimension in the study of graphene is that it can be incorporated
into a field effect device\cite{geimnatmat} and doped away from half filling by changing
the gate voltage. This means that the position of the chemical potential
can be varied with respect to the Dirac point where conical valence and
conduction band meet at two specific $K$ points 
in the Brillouin zone. As is the
case in ordinary metals the electron-phonon interaction\cite{Grimvall,Prange}
 (EPI)  renormalizes
the bare bands and changes the properties of graphene. Here we consider
specifically the electronic density of states (DOS). In metals, where the DOS is
essentially energy independent on the scale of phonon energies,
 a well-known
result is that the electron-phonon self-energy entirely drops out of the 
problem 
and the DOS retains its bare value. In graphene, as we will show in
this paper, this is no longer the case and there is an imprint of the 
EPI in the renormalized DOS. This is expected in systems where the DOS
varies on an energy scale comparable to phonon energies.\cite{boza}
 It also arises in
finite band systems, with the top and the 
bottom of the band are modified in a particularly important way
by interactions with the phonons.\cite{frank,cappelluti,anton,Engelsberg}

Another known result in simple metals is that
 the EPI renormalizations\cite{Grimvall,Prange,Mori} drop out of the
DC value of the conductivity but significantly renormalize its AC value
and phonon-assisted Holstein processes become possible. For 
graphene the DC conductivity 
which is due to both intra and interband 
transitions  is also unaffected by the EPI. However, the intraband AC 
Drude contribution
has its optical spectral weight reduced by a factor of ($1+\lambda^{\rm eff}$),
where $\lambda^{\rm eff}$ is the effective mass renormalization parameter
(dependent on value of the chemical potential), and its scattering rate is 
also reduced by the same factor. Optical spectral weight is shifted to phonon-assisted
Holstein side bands originating from the incoherent part of the electron spectral
density. The interband transitions, which are strongly
suppressed by Pauli blocking below twice the renormalized chemical 
potential, also exhibit phonon-assisted side bands which add on to the
intraband Holstein processes at photon energies above the phonon energy.
The universal value of the background conductivity at higher energies
is suppressed below its bare band value $\sigma_0=\pi e^2/2h$.

In Section II, we introduce the model of Park et al.\cite{Parks,parknano,park}
for the electron-phonon interaction in graphene which we write in a 
more general form
so as to include the self-consistent electronic density of states (DOS)
in the self-energy itself. Iterations of the self-energy to obtain a self-consistent
value of the DOS is known to be important in finite bands particularly in the
region of the band edge. The electron-phonon interaction renormalizes
the electronic motion at the Fermi surface through a renormalization factor
($1+\lambda^{\rm eff}$).\cite{julesrmp} In addition, the chemical potential is changed from its
bare band value as is the relative position of the Dirac point. In the limit of
small chemical potential, analytic formulas are obtained for these
renormalizations and are compared with more general numerical
results. Our results for the renormalized DOS show phonon signatures
at small energies and a renormalization of the band edge with renormalized
bands no longer ending abruptly as in the bare band. An analytic 
formula is derived for the modification of the Dirac point due to interactions.

In Section III, we consider more specifically phonon structure in the renormalized
DOS, a topic on which 
we have some preliminary and complementary work.\cite{sgrph}  
To this end, we introduce a Lorentzian model for the phonon distribution
instead of the simple Einstein model of Park et al.\cite{Parks} This provides a more
realistic description of the structure that is expected to be seen in
experiment.\cite{li} To see this structure more clearly, 
we take a first derivative
and find that the Lorentzian line shape of our assumed boson spectrum is
faithfully reproduced in these plots except for a small shift in energy.
Section IV deals with aspects of quasiparticle renormalization as they
show up in angle resolved photoemission spectra (ARPES).\cite{calandra,bostwick,dassarma,polini}
 We particularly
emphasize the shift in chemical potential\cite{luttinger} revealed in such curves and note
that electron-phonon renormalizations remain to high energies. Section V
deals with optical conductivity\cite{peres,ando,falkovsky,paper1,paper2,paper3,gusynin,dpaper,wang,pedersen} starting  
with the effect of the electron-phonon renormalization on the low
frequency conductivity. We begin with the DC conductivity which is found
to be unrenormalized. Conditions for the observation of the universal
limit are discussed and the contribution to this value of intra and interband
transitions are separately determined. The rise of the conductivity $\sigma(\Omega)$
from its universal DC to its universal AC background
value\cite{gusynin} with increasing frequency $\Omega$
is described. Analytic formulas are provided for the electron-phonon renormalization that arises for finite doping when $\mu$ is much larger than the
impurity scattering rate $1/\tau$. We find that in this case,
the intraband transitions provide a Drude-like contribution to the conductivity
which, just as is known to be the case in ordinary metals,\cite{Mori} is renormalized
by the mass enhancement factor ($1+\lambda^{\rm eff}$) in two ways. The
effective plasma frequency of the Drude as compared to its free band case is
reduced by $1/(1+\lambda^{\rm eff})$ and the scattering rate $1/\tau$
also goes into $(1/\tau)/(1+\lambda^{\rm eff})$. In addition, optical
spectral weight is 
transferred into Holstein phonon-assisted absorption sidebands. On
the other hand and in sharp contrast, the interband transitions at small
$\Omega$ remain unrenormalized but beyond the phonon energy also exhibit
Holstein sidebands. In section VII, the infrared region of the 
conductivity\cite{StauberGeim,Kuzmenko,Nair,katsnelson}
is considered more explicitly with the specific aim of understanding
separately the role played by intra and interband transitions. It is
explained how interband transitions which would be Pauli blocked in the pure
bare
band case can nevertheless take place in the interacting system although
with reduced spectral weight. This arises because in the interacting
system Bloch states are never occupied with probability one and so
Pauli blocking is not complete. The partial sum rule on the optical
conductivity is discussed. Section VIII contains a summary and conclusions.

\section{Theoretical Background}

In graphene, the massless Dirac nature of the electrons means that their 
energy ($\epsilon_{\bf k}$) is linear in momentum ($\hbar {\bf k}$)
and consists of two branches, i.e.,
$\epsilon_{\bf k}=\pm\hbar v_0|{\bf k}|$
with the velocity $v_0$ equal to about $10^6$ m/sec and ${\bf k}$ the
wave vector measured from one of the two $K$ points in the Brillouin
zone where valence and conduction bands meet. The $\pm$ corresponds
to the upper and lower Dirac cones or particles and holes, respectively. Here,
$\epsilon$ can serve as the label for the absolute value of momentum. 
The carrier spectral
density $A(\epsilon,\omega)$ can be written in terms of the self-energy
$\Sigma(\omega)$ as
\begin{equation}
A(\epsilon,\omega)=\frac{1}{\pi}\frac{-{\rm Im}\Sigma(\omega)}
{[\omega-{\rm Re}\Sigma(\omega)+\mu-\epsilon]^2+[{\rm Im}\Sigma(\omega)]^2}.
\label{eq:spectral}
\end{equation}
Valence and conduction cones are both part of Eq.~(\ref{eq:spectral})
and correspond, respectively, to negative and positive $\epsilon$ label.
As defined, $\Sigma(\omega)$ is the self-energy such that $\omega=0$
corresponds to the Fermi level and $\mu$ is the chemical potential
of the interacting system. 
The electron-phonon interaction in graphene has been discussed by many
authors\cite{Parks,calandra,dassarma} including recent work\cite{paper1,paper2,paper3}
directed towards understanding its effect on the optical conductivity.
Here we find it convenient to follow the formulation of C.H. Park et al.
\cite{Parks}
later also generalized to bilayers and graphite.\cite{parknano}
These authors proceed within an ab initio pseudopotential density functional
calculation of the electronic bands in the local density approximation.
They further calculate the phonon frequencies and polarization as well
as the electron-phonon matrix elements from which they construct the electron
self-energy. The importance of this work to the present discussion is that
in the end they provide a simplified model for the self-energy
which they find captures all essential elements of their sophisticated numerical 
calculations. They show that to a good approximation, one can think of the electrons
as coupled to a single phonon of frequency $\omega_E=200$ meV and provide
a value for the relevant coupling. The self-energy is independent of electron
momentum label as well as band index, electrons or holes, a simplification
they trace\cite{parknano} to the Dirac nature of the electronic states in
graphene. Based on these simplifying ideas, we begin with a self-energy
for temperature $T=0$ of the form\cite{frank,cappelluti}
\begin{widetext}
\begin{equation}
\Sigma(\omega)=\int_{-\infty}^{+\infty}d\omega'\frac{N(\omega')}{N_\circ}
\frac{A}{W_C}
\biggl[\frac{\theta(\omega')}{\omega-\omega'-\omega_E+i0^+}
+\frac{\theta(-\omega')}{\omega-\omega'+\omega_E+i0^+}\biggr] .
\label{eq:Sigma}
\end{equation}
\end{widetext}
In Eq.~(\ref{eq:Sigma}),
$\omega_E$ is the Einstein oscillator energy, $\omega$ is in units of energy,
$A$ is the coupling, $\theta(\omega)$ is 
the Heaviside function,
and $N_\circ = 2/\pi\hbar^2v_0^2$. 
 $W_C$ is the cut off on the Dirac cone given by
$\sqrt{\pi\sqrt{3}}t$, with $t$ the nearest neighbor hopping parameter
taken to be 3 eV which corresponds to
$W_C\simeq 7000$ meV. In Eq.~(\ref{eq:Sigma}), we have
added an additional element not included in the previous work of C.H. Park et al.\cite{Parks}
but which can be important in the case of graphene: $N(\omega)$
is the {\it self-consistent} electronic density of states given by
\begin{equation}
\frac{N(\omega)}{N_\circ}=\int_{-W_C}^{W_C}d\epsilon \frac{|\epsilon|}{\pi}
\frac{-{\rm Im}\Sigma(\omega)}
{[\omega-{\rm Re}\Sigma(\omega)+\mu-\epsilon]^2+[{\rm Im}\Sigma(\omega)]^2}.
\label{eq:dos}
\end{equation}
This can be evaluated to give
\begin{widetext}
\begin{equation}
\frac{N(\omega)}{N_\circ}=-\frac{\tilde\omega}{\pi }\biggl[
\arctan\biggl(\frac{\tilde\omega-W_C}{\Gamma}\biggr)
+\arctan\biggl(\frac{\tilde\omega+W_C}{\Gamma}\biggr)
-2\arctan\biggl(\frac{\tilde\omega}{\Gamma}\biggr)\biggr]
\displaystyle
+\frac{\Gamma}{2\pi}\ln\biggl(\frac{[(\tilde\omega-W_C)^2+\Gamma^2]
[(\tilde\omega+W_C)^2+\Gamma^2]}{(\tilde\omega^2+\Gamma^2)^2}\biggr),
\label{eq:dosformula}
\end{equation}
\end{widetext}
where $\Gamma=-{\rm Im}\Sigma(\omega)$ and $\tilde\omega=\omega-{\rm Re}\Sigma(\omega)+\mu$.
In good metals, the electronic density of states (DOS) in the energy range
important for the electron-phonon interaction around the Fermi energy, does
not vary significantly. Only its value at the chemical potential is relevant.
In such systems,
the linear in $\epsilon$ factor $|\epsilon|$ is not present in Eq.~(\ref{eq:dos}) and extending the integration limits to infinity gives a constant independent
of ${\rm Im}\Sigma(\omega)$ and ${\rm Re}\Sigma(\omega)$.
In more complicated metals such as the A15 compounds, as an example, it was
recognized early on that such an approximation is no longer valid\cite{boza}
and in this instance it is the self-consistent DOS\cite{frank,cappelluti,anton}
that enters Eq.~(\ref{eq:Sigma}), as in the work of Engelsberg and
Schrieffer.\cite{Engelsberg} This is also relevant for finite bands.\cite{frank,cappelluti,anton}
For graphene, the bare band structure density of states
is linear in energy rather than constant
and hence we expect significant changes could be introduced
into $\Sigma(\omega)$ when equations (\ref{eq:Sigma}) and (\ref{eq:dos})
are iterated to convergence. We will test this in what follows.

\begin{figure}[ht]
\begin{picture}(250,200)
\leavevmode\centering\includegraphics{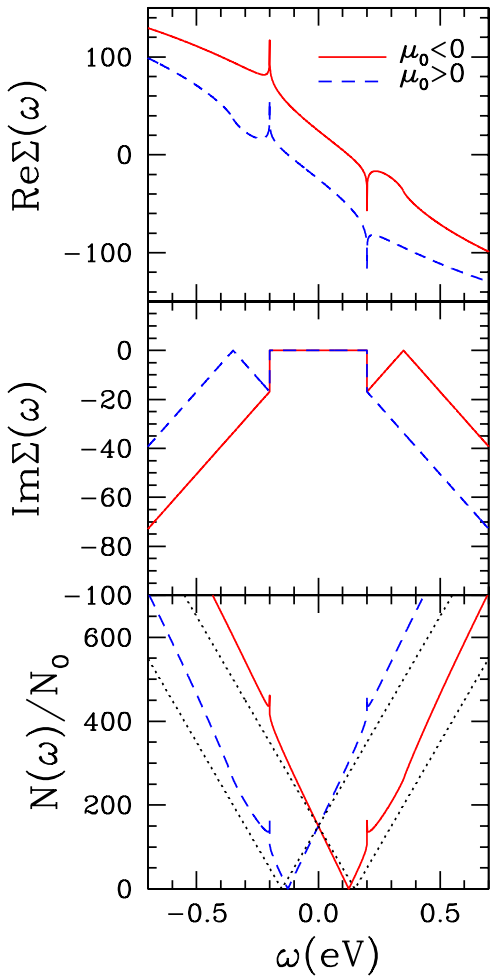}
\end{picture}
\vskip 160pt
\caption{(Color online) Real (top) and imaginary (middle) part of the self-energy
$\Sigma(\omega)$ and the density of states $N(\omega)/N_\circ$ (bottom) (all
in units of meV) 
as a function of $\omega$ in eV for $|\mu_0|=$150 meV.
Dashed is for $\mu_0>0$ and solid is for $\mu_0<0$. The bare band DOS is indicated
by the dotted curve.
}
\label{Fig1}
\end{figure}

As a first approximation, we can replace $N(\omega)$ in Eq.~(\ref{eq:Sigma})
by its bare non-interacting value given by
\begin{equation}
\frac{N(\omega)}{N_\circ}=
\begin{cases} |\omega+\mu_0|, & {\rm for} -W_C-\mu_0<\omega<W_C-\mu_0 ,\\
           0, & {\rm otherwise,}
\end{cases}
\label{eq:baredos}
\end{equation}
where $\mu_0$ is the chemical potential for the bare bands, i.e. without
electron-phonon renormalization. All states with $\omega <0$ are occupied
and with $\omega>0$ unoccupied, so that $\omega=0$ sets the boundary 
between occupied and unoccupied energies and by choice
this remains the case even
when interactions are included. Note that the value of $\mu_0$ sets the
doping $\rho$ (per unit area) with 
$\rho={\rm sgn}(\mu_0)\mu_0^2/(\pi\hbar^2 v_0^2)$, with $\mu_0$ negative
for holes and positive for electrons. This is illustrated in the bottom
frame of Fig.~\ref{Fig1} by the black dotted lines for the case of
$|\mu_0|=150$ meV. The Dirac point is at $-150$ meV ($+150$ meV) for 
electron (hole) doping and corresponds to zero DOS or $\omega=-\mu_0$ 
from Eq.~(\ref{eq:baredos}). For this bare $N(\omega)$ the self-energy
of Eq.~(\ref{eq:Sigma}) can be evaluated analytically but is different for 
$\mu_0>0$ and $\mu_0<0$. 
For $\mu_0>0$ (writing $\Sigma_{\mu_0>0}(\omega)$)
\begin{widetext}
\begin{eqnarray}
{\rm Re}\Sigma_{\mu_0>0}(\omega)&=&\displaystyle
\frac{A}{W_C}\Biggl\{\omega_E \ln\Biggl|
\displaystyle
\frac{(W_C+\omega_E-\omega-\mu_0)(\mu_0+\omega+\omega_E)^2}{(\omega^2-\omega_E^2)
(W_C+\omega+\omega_E+\mu_0)}\Biggr|\nonumber\\
&&\quad -(\mu_0+\omega)\ln\Biggl|
\displaystyle
\frac{(W_C+\omega_E-\omega-\mu_0)(W_C+\omega+\omega_E+\mu_0)(\omega+\omega_E)}
{(\omega-\omega_E)(\omega+\mu_0+\omega_E)^2}\biggr|\Biggr\}
\label{eq:Sigmamugt}
\end{eqnarray}
\end{widetext}
and for negative $\mu_0$
\begin{equation}
{\rm Re}\Sigma_{\mu_0<0}(\omega)=-{\rm Re}\Sigma_{|\mu_0|}(-\omega) .
\label{eq:resigma}
\end{equation}
This symmetry between positive and negative values of chemical potential also holds
in the case when the full self-consistent density of states (DOS) $N(\omega')$ is used in Eq.~(\ref{eq:Sigma}). Before showing this explicitly,
we note that 
\begin{widetext}
\begin{equation}
-{\rm Im}\Sigma(\omega)=
\begin{cases}\displaystyle\frac{\pi A}{W_C}|\omega-\omega_E+\mu_0|, & {\rm for}\quad \omega_E<\omega<W_C-\mu_0+\omega_E ,\\
           \displaystyle\frac{\pi A}{W_C}|\omega+\omega_E+\mu_0|, & {\rm for}\quad -\omega_E>\omega>-W_C-\mu_0-\omega_E ,
\end{cases}
\label{eq:imsigmafull}
\end{equation}
\end{widetext}
and zero outside these intervals. For negative values of $\mu_0$ we note
\begin{equation}
{\rm Im}\Sigma_{\mu_0<0}(\omega)={\rm Im}\Sigma_{|\mu_0|}(-\omega) .
\label{eq:imsigma}
\end{equation}
Using Eqs.~(\ref{eq:resigma}) and (\ref{eq:imsigma}) in Eq.~(\ref{eq:dos})
for the self-consistent DOS, we see that 
\begin{equation}
N_{\mu_0<0}(\omega)=N_{|\mu_0|}(-\omega)
\end{equation}
and this relationship implies that the symmetries of  Eqs.~(\ref{eq:resigma}) and (\ref{eq:imsigma}) apply even when the full self-consistent DOS is
used in Eq.~(\ref{eq:Sigma}) for the self-energy.

In Fig.~\ref{Fig1}, we show results for ${\rm Re}\Sigma(\omega)$ (top frame),
${\rm Im}\Sigma(\omega)$ (middle frame) and $N(\omega)/N_\circ$ (bottom
frame), where $N(\omega)$ is a first iteration of Eq.~(\ref{eq:Sigma})
which should be sufficient for small enough electron-phonon interaction. The dashed (blue) lines
are for $\mu_0>0$ (electron doping)
and the solid (red) for $\mu_0<0$ (hole doping). The dotted (black) 
line in the
bottom frame is the bare band density of states shown for comparison.
Here the chemical potential $\mu_0$ was taken to be $|\mu_0|=150$~meV, 
$W_C=7000$~meV and
$\omega_E$ was taken to be 200~meV 
as in Ref.~\cite{Parks}. The value of the electron-phonon
$A$ parameter was set at 250~meV.  The corresponding spectral mass enhancement
$\lambda\equiv 2A/\omega_E=2.5$. As we will show, 
this parameter is very different from
the true mass enhancement which we will call $\lambda^{\rm eff}$ and which
describes for graphene the renormalization of the velocity because the carriers
are massless.

Note that the symmetry incorporated into Eq.~(\ref{eq:resigma}) is satisfied
in the top frame for ${\rm Re}\Sigma(\omega)$. We see logarithmic-type
 singularities\cite{frank}
at $\omega=\pm\omega_E$ but these would be smeared if one had used a distributed
phonon spectra rather than an Einstein model. The logarithmic singularity can be traced
to a factor $(\omega\pm\omega_E)$ in 
the argument of the logarithm in
Eq.~(\ref{eq:Sigmamugt}) and these will lead 
to sharp structure in the electronic density of states. For later reference, we
note an important exception. For zero chemical potential $\mu_0=0$, cancellations
occur in Eq.~(\ref{eq:Sigmamugt}) and the singularities are removed.

${\rm Re}\Sigma(\omega=0)$
is not zero but $-24.7$ ($+24.7$) meV for $\mu_0>0$  ($\mu_0<0$). This quantity
relates interacting ($\mu$) and noninteracting ($\mu_0$) chemical potential\cite{boza,Engelsberg,luttinger} with 
\begin{equation}
\mu=\mu_0+{\rm Re}\Sigma(\omega=0).
\label{eq:interactmu}
\end{equation}
If for simplicity we assume all three energy scales $\omega$, $\mu$,
and $\omega_E$ to be small as compared with the large band cut off $W_C$,
then 
Eq.~(\ref{eq:Sigmamugt}) for ${\rm Re}\Sigma(\omega)$ gives at $\omega=0$ ($\mu_0>0$)
\begin{equation}
{\rm Re}\Sigma(\omega=0)=\frac{2A}{W_C}\biggl\{\omega_E\ln\biggl|\frac{\mu_0+
\omega_E}{\omega_E}\biggr|-\mu_0\ln\biggl|\frac{W_C}{\mu_0+\omega_E}\biggr|\biggr\}.
\end{equation}
The shift between interacting and noninteracting chemical potential
is zero only for the case $\mu_0=0$ (particle-hole symmetry). For $\mu_0\ll\omega_E$
\begin{equation}
{\rm Re}\Sigma(\omega=0)\simeq -\frac{2A}{W_C}\mu_0
\biggl[\ln\biggl(\frac{W_C}{\omega_E}\biggr)-1\biggr] .
\label{eq:13}
\end{equation}
In Eq.~(\ref{eq:13}), $W_C$ is involved, the limit $W_C\to\infty$
cannot be taken, and ${\rm Re}\Sigma(\omega=0)$ is a fraction
of $\mu_0$ because $A/W_C$ is a small number. In terms of the
spectral $\lambda$ introduced earlier ($\lambda=2A/\omega_E$), the 
coefficient $2A/W_C=\lambda(\omega_E/W_C)$ with $\omega_E/W_C\sim 1/35$
for realistic parameters corresponding  to graphene.

For small but finite values of $\omega$, we get from Eq.~(\ref{eq:Sigmamugt}),
in the same limits, ($|\mu_0|<\omega_E$),
\begin{equation}
{\rm Re}\Sigma(\omega)\simeq -\frac{2A}{W_C}\omega
\biggl[\ln\biggl|\frac{W_C}{\mu_0+\omega_E}\biggr|-\biggl(1-\frac{|\mu_0|}{\omega_E}
\biggr)\biggr]+{\rm Re}\Sigma(\omega=0).
\end{equation}
Denoting the carrier effective mass renormalization due to the 
electron-phonon interaction by $\lambda^{\rm eff}$ then for $\omega\to 0$,
the self-energy is given by 
\begin{equation}
{\rm Re}\Sigma(\omega)= -\lambda^{\rm eff}\omega +{\rm Re}\Sigma(\omega=0).
\label{eq:15}
\end{equation}
The renormalized energy $(E_k)$ as measured, for example, in angle-resolved
photoemission spectroscopy (ARPES) is related to the bare energy near the
Fermi energy by
\begin{equation}
E_k=\frac{\epsilon_k-\epsilon_{k_F}}{1+\lambda^{\rm eff}}
=\frac{\pm\hbar v_0(k-k_F)}{1+\lambda^{\rm eff}} ,
\label{eq:16}
\end{equation}
with $\hbar k_F$, the Fermi momentum of the bare band,
so $\mu_0=\pm\hbar v_0k_F$, with the $+$ sign for electron doping
and $-$ for holes.  We see that interactions renormalize the 
bare effective speed of light
 $v_0$ to $v^*_0=v_0/(1+\lambda^{\rm eff})$. To see this, we
return to the spectral density  of Eq.~(\ref{eq:spectral}) and note that 
bare and renormalized energies near the Fermi surface are obtained when the
first bracket in the denominator is set equal to zero:
\begin{equation}
E_k+\lambda^{\rm eff}E_k-{\rm Re}\Sigma(\omega=0)+\mu=\epsilon_k ,
\end{equation}
which gives Eq.~(\ref{eq:16}) 
with $(k-k_F)$ small.
For $\mu_0\ll\omega_E$ we get
\begin{equation}
\lambda^{\rm eff}=\frac{2A}{W_C}\biggl[\ln\biggl|\frac{W_C}{\omega_E}\biggr|-1
\biggr],
\label{eq:17}
\end{equation}
which is $\sim 0.19$ for the parameters of Fig.~\ref{Fig1}. This value is
larger than calculated in density functional theory\cite{Parks} but
smaller than measured experimentally\cite{bostwick}.
Note that in terms of Eq.~(\ref{eq:17}), Eq.~(\ref{eq:13}) can be written
as 
\begin{equation}
{\rm Re} \Sigma(\omega=0)=-\mu_0\lambda^{\rm eff}
\end{equation}
and so noting Eq.~(\ref{eq:interactmu})
\begin{equation}
\mu=\mu_0(1-\lambda^{\rm eff})\simeq\frac{\mu_0}{1+\lambda^{\rm eff}}
\label{eq:21}
\end{equation}
since $\lambda^{\rm eff}$ is small so that in this simple limit the factor
$(1+\lambda^{\rm eff})$ also renormalizes the value of the bare
chemical potential.

The middle frame of Fig.~\ref{Fig1} gives first iteration results for the
imaginary part (${\rm Im}\Sigma(\omega)$) of the self-energy which clearly
satisfy the symmetry of Eq.~(\ref{eq:imsigma}). ${\rm Im}\Sigma(\omega)$
is zero between $-\omega_E<\omega <\omega_E$ at which point it jumps up to
a finite value because only then can a quasiparticle decay by
boson emission. In an ordinary metal the scattering would remain
constant above $\omega_E$ because the final density of electronic
states 
at $\omega-\omega_E$ is
constant. 
This is not true for graphene
where $N(\omega-\omega_E)$ varies and
 the scattering rate reflects the energy dependence
of the underlying bare band structure seen in the lower frame
of Fig.~\ref{Fig1} as the black dotted curve. Looking at the red solid
curve for energy less than $\omega_E$, the $|{\rm Im}\Sigma(\omega)|$
increases linearly with $\omega$ as does $N_0(\omega)$ and for
$\omega>\omega_E$ we see first a drop towards the zero in $|{\rm Im}\Sigma(\omega)|$
 after
which it too increases linearly, imaging the bare density of states
as captured in Eq.~(\ref{eq:imsigmafull}). 
In more
realistic calculations there would be a nonzero contribution to the self-energy
from acoustic phonons and ${\rm Im}\Sigma(\omega)$ would never be exactly
zero as we have it here. But from the work of Ref.~\cite{Parks}, this
is small in graphene. Finally, we note that for a fixed value of $\pm\mu_0$,
${\rm Im}\Sigma(\omega)$ is not symmetric with respect to $\omega\to -\omega$
as it would be in good metals. Here such a symmetry results only when 
$\mu_0=0$, i.e. the case of no doping.

In the lower frame of Fig.~\ref{Fig1}, we show our one iteration
results for the DOS $N(\omega)$ (solid red curve for 
$\mu_0<0$ and dashed blue for $\mu_0>0$) and compare with the bare 
case (dotted black curves). $N(\omega)$ is not symmetric about $\omega=0$ and
the renormalizations are not the same for electron and hole branches. Also,
the Dirac point is shifted to higher energy as compared to its bare band
position for $\mu_0>0$ and to lower energy for $\mu_0<0$. The shift
becomes greater with increased doping. Another feature
to notice is that bare and renormalized DOS have exactly the same value at
$\omega=0$, which is where by choice we have taken the  Fermi level to be
in both cases. This makes sense because the imaginary part of $\Sigma(\omega)$ at
$\omega=0$ is always zero and so the Lorentzian forms in Eq.~(\ref{eq:dos})
reduce to delta functions and for $\mu_0>0$, as an example,
\begin{equation}
\frac{N(\omega=0)}{N_\circ}=\mu-{\rm Re}\Sigma(\omega=0)=\mu_0,
\end{equation}
which is the value of the bare density of states at the
bare chemical potential.
The value of the DOS at the Fermi surface remains
pinned to its noninteracting value. Another important feature of our
DOS results is that phonon structures are clearly seen as small kinks in the
curve. This will be smeared somewhat when a distributed spectrum is used in
computing the self-energy rather than the Einstein mode of Eq.~(\ref{eq:Sigma}).
We will return to this issue in a later section where we focus more 
specifically on phonon structures. 
For the moment, we note that our results for $N(\omega)$ versus $\omega$
are in striking contrast to what is found to 
apply in conventional metals. In that case, the density of electronics
states around the Fermi surface is constant on the scale of a few
times the phonon energy and no renormalization of the DOS due to the electron-phonon
interaction is expected or observed. Further in conventional metals, the 
energy bands are rigidly filled as the chemical potential is varied.
For graphene, when coupling to a boson spectrum is included, the bands become 
renormalized differently for {\it each value} of the chemical
potential $\mu_0$.
These are not {\it rigid} bands.

\begin{figure}[ht]
\begin{picture}(250,200)
\leavevmode\centering\includegraphics{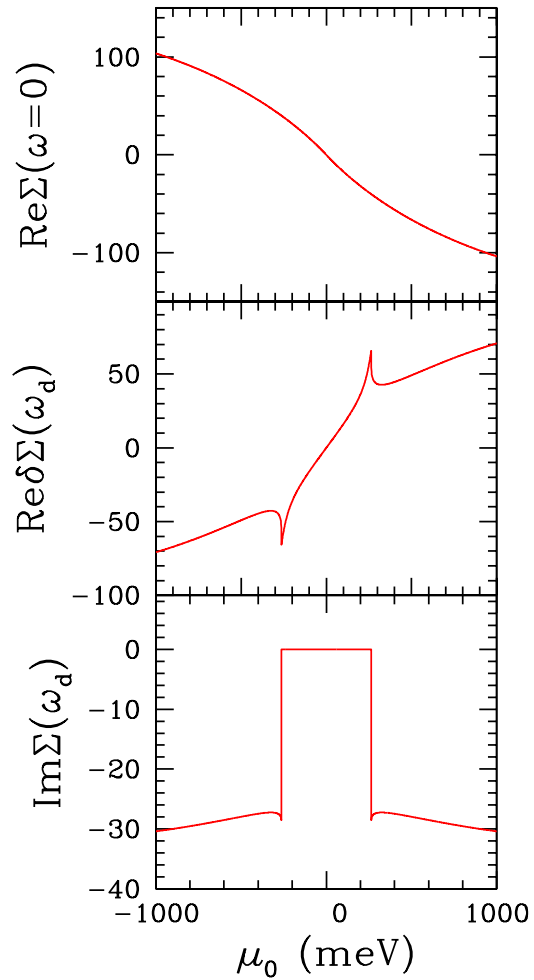}
\end{picture}
\vskip 160pt
\caption{(Color online) Results of a first iteration for the shift in 
chemical potential ${\rm Re}\Sigma(\omega=0)$ (top frame), the shift in
position of the Dirac point ${\rm Re}\delta\Sigma(\omega_d)$ (middle frame)
and the imaginary part of the self-energy ${\rm Im}\Sigma(\omega_d)$
at the Dirac point (lower frame). All of these quantities are shown in 
units of meV.
}
\label{Fig2}
\end{figure}

Next we look at how the position of the Dirac point shifts from its bare
value at $\omega=-\mu_0$ when interactions are included.
In the interacting system, the Dirac point can still be identified with
zero momentum $k=0$ (momentum label) which corresponds to $\epsilon=0$
(energy label) as $\epsilon=\pm\hbar v_0|{\bf k}|$. If there was no broadening (but of course
there is), the spectral function would become infinite (peaks) at 
$\omega-{\rm Re}\Sigma(\omega)+\mu=0$. If we write
\begin{equation}
{\rm Re}\Sigma(\omega)={\rm Re}\Sigma(\omega=0)+{\rm Re}\delta\Sigma(\omega)
\label{eq:sigmadelta}
\end{equation}
where ${\rm Re}\delta\Sigma(\omega)=0$ at $\omega=0$ by arrangement,
we get
\begin{equation}
\omega-{\rm Re}\Sigma(\omega=0)+\mu-{\rm Re}\delta\Sigma(\omega)=0
\label{eq:diracsolve}
\end{equation}
as the transcendental equation which needs to be solved to
obtain the energy of the Dirac point in the interacting band. For one
iteration, ${\rm Re} \Sigma(\omega)$ is given by Eq.~(\ref{eq:Sigmamugt})
from which ${\rm Re}\delta\Sigma(\omega)$ can easily be extracted.
Noting Eq.~(\ref{eq:interactmu}), Eq.~(\ref{eq:diracsolve})
simplifies to 
\begin{equation}
\omega_d=-\mu_0+{\rm Re}\delta\Sigma(\omega_d) ,
\label{eq:diracsimple}
\end{equation}
where $\omega_d$ is the energy of the Dirac point in the interacting
system. For the bare band case ${\rm Re}\delta\Sigma(\omega_d)$ in
Eq.~(\ref{eq:diracsimple}) is zero and we recover the known result that
the Dirac point is at $-\mu_0$. ${\rm Re}\delta\Sigma(\omega_d)$ which we
plot as a function of $\mu_0$ in the middle frame of Fig.~\ref{Fig2} gives the shift in
$\omega_d$ away from $-\mu_0$ resulting from the electron-phonon interaction.
It has the same sign as $\mu_0$ and hence displaces the Dirac point to the
right of its bare value for $\mu_0>0$ and to the left for $\mu_0<0$ as we noted
before. In the upper frame of Fig.~\ref{Fig2} we show a related quantity
namely ${\rm Re}\Sigma(\omega=0)$ which is the shift in chemical potential
($\mu$) from its bare band value ($\mu_0$). It carried the opposite sign to
$\mu_0$ and so shifts $\mu$ to the left of $\mu_0$ for positive $\mu_0$
and vice versa for negative $\mu_0$. This shift in chemical potential
is expected because the interactions modify the density of states and hence
to keep the carrier imbalance fixed, the position of the Fermi level needs
to be modified. The question of how this shift might manifest itself
will be addressed in a later section.
Shifts in the
Dirac point position can be seen in ARPES\cite{bostwick} and in scanning tunneling
microscopy (STM)\cite{zhang}. Noting that the electron-phonon
interaction makes only a small contribution to the measured total quasiparticle
scattering rates\cite{park} seen in graphene and that the electron-electron
interactions are needed to understand these, i.e., 
particle-hole and plasmons,\cite{park} any conclusive comparison with experiment would need to include these
additional interactions and so we do not attempt here a comparison with data.
Before leaving this section, we note that for $\mu_0\ll\omega_E$ we can show
that 
\begin{equation}
\omega_d\simeq -\frac{\mu_0}{1+\lambda^{\rm eff}}
\label{eq:26}
\end{equation}
from (\ref{eq:diracsimple}) with ${\rm Re}\delta\Sigma(\omega_d)$ replaced by
${\rm Re}\delta\Sigma(-\mu_0)$. Again,
to a first approximation, the $(1+\lambda^{\rm eff})$ factor
renormalizes  the position in energy of the Dirac
point. In this approximation, the magnitude of $\omega_d$, i.e., $|\omega_d|$,
is equal to the magnitude of the renormalized chemical potential
$|\mu|$. Except for
prominent phonon structures seen in the middle of Fig~\ref{Fig2}, not present
in the top frame, $|{\rm Re}\delta\Sigma(\omega_d)|$ and $|{\rm Re}\Sigma(\omega=0)|$
are not very different in magnitude as compared with the magnitude of
$\mu$
itself 
and so  renormalized values of $|\omega_d|$ and $|\mu|$ are close to each other,
even for a general value of chemical potential.

\begin{figure}[ht]
\begin{picture}(250,200)
\leavevmode\centering\includegraphics{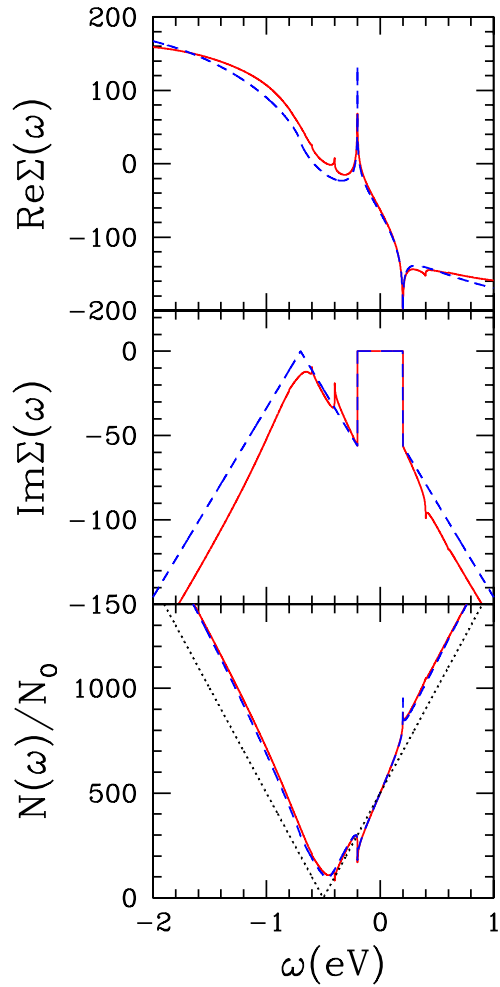}
\end{picture}
\vskip 160pt
\caption{(Color online) Same quantities as for Fig.~\ref{Fig1} 
but now the
self-energy and density of states have been self-consistently iterated
and $\mu_0=500$ meV which is greater than $\omega_E$.
The solid curves are for the iterated case and the dashed curves are
for the initial uniterated results. The dotted
curve is the bare density of states. All quantities on the
y-axis are in meV.
}
\label{Fig3}
\end{figure}

\begin{figure}[ht]
\begin{picture}(250,200)
\leavevmode\centering\includegraphics{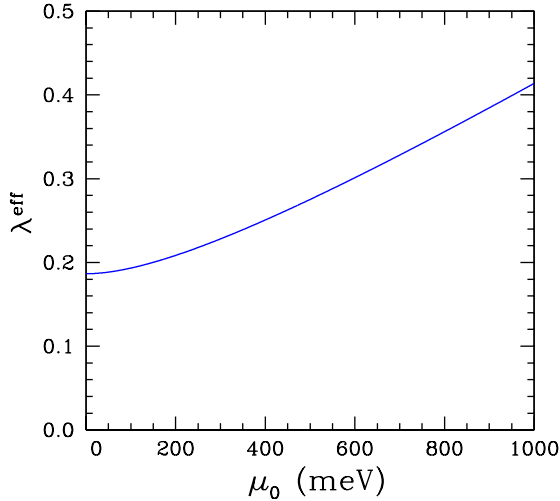}
\end{picture}
\vskip 20pt
\caption{(Color online)  The mass renormalization parameter $\lambda^{\rm eff}$
at the Fermi level as a function of bare chemical potential $\mu_0$. 
}
\label{Fig4}
\end{figure}

In Fig.~\ref{Fig3}, we consider the case of $\mu_0=500$ meV
and compare our initial step uniterated results shown
as the dashed lines with results obtained  (solid curves) when DOS (Eq.~(\ref{eq:dos}))
and self-energy (Eq.~(\ref{eq:Sigma})) are iterated to convergence.
We see some changes in real and imaginary part of $\Sigma(\omega)$
top and middle frames, respectively. We note however that the change in slope
of the ${\rm Re}\Sigma(\omega)$ out of zero is not changed much in
the solid curve which means that the one iteration estimate of the 
effective mass renormalization $\lambda^{\rm eff}$ 
is quite good. Results for 
$\lambda^{\rm eff}$ are shown in Fig.~\ref{Fig4} as a function of
$\mu_0$.
Its value increases by a factor of two over the range shown.

\begin{figure}[ht]
\begin{picture}(250,200)
\leavevmode\centering\includegraphics{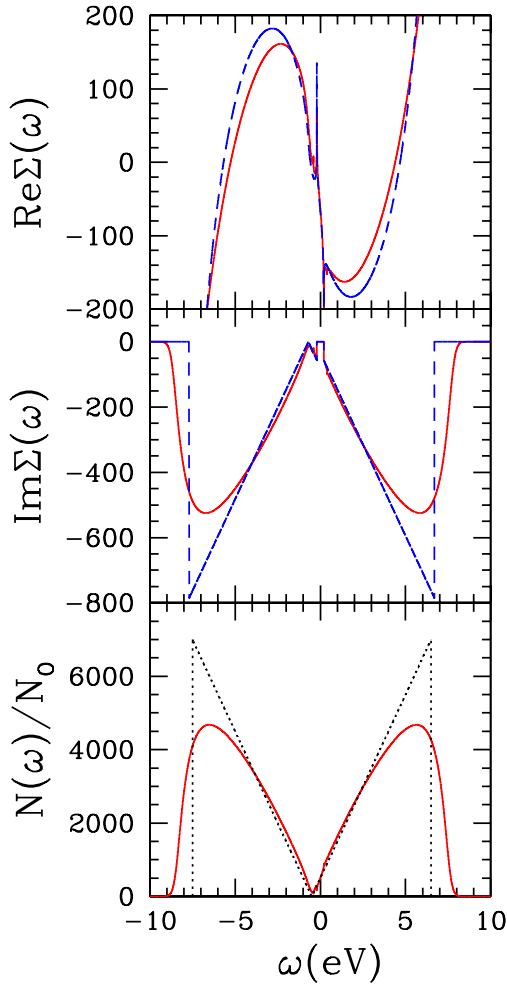}
\end{picture}
\vskip 170pt
\caption{(Color online) Same as for Fig.~\ref{Fig3} but shown over
a larger range of $\omega$ spanning the band edge. 
}
\label{Fig5}
\end{figure}

The differences between iterated and noniterated results for the 
$-{\rm Im}\Sigma(\omega)$ shown in the middle frame of Fig.~\ref{Fig3} are larger and can become
significant.
It is clear that self-consistency could be important in some quantitative
estimation of electron-phonon effects on the quasiparticle
scattering rate and that LDA can underestimate these
in some cases.
There are
also changes in the resulting DOS shown in the bottom frame but these are
small in the case shown. Effects of self-consistency become
essential, however, when the band edge is considered.
Fig.~\ref{Fig5} is the same as Fig.~\ref{Fig3}
but now the entire band width is shown within the linearized Dirac cone 
approximation with cut off $W_C$. On this scale the phonon structures
visible as sharp peaks in Fig.~\ref{Fig3} are hardly seen. Note that in the
 one iteration case the imaginary part of $\Sigma(\omega)$ (middle frame) dashed curve
becomes zero at the bare band edge because the DOS runs out at this point
but the solid curve which results when we iterate remains finite outside this
range before eventually going to zero. These features of the renormalization
get reflected in the density of state shown in the lower frame of Fig.~\ref{Fig5}
in which we have also added for reference the bare DOS $N_0(\omega)$
as the dotted (black) curve. It is clear that the electron-phonon interaction
profoundly renormalizes the band edge. These effects have been studied before
for other models of finite bands in, for example, Refs.~\cite{frank} and \cite{anton}.
To obtain a good characterization of the renormalized band profile (red curve) around
the bare band edge $W_C$, it is essential to iterate.\cite{frank,anton}
We note that the top of the renormalized band extends to higher energies
as compared with the bare band and the bottom extends to lower energies.
It is the phonon energy which sets the scale for this smearing beyond the bare
band edge. 
 Of course
the total number of states must be preserved, i.e. the area under $N(\omega)/N_\circ$
is the same for bare (dotted black) and renormalized (solid red) curves.
Due to interactions, i.e. damping, the states below $W_C$ get depleted and
so must reappear at higher energies leading to an increase in the effective
band width. We have verified that to within our numerical accuracy the sum rule
on $N(\omega)/N_\circ$ was indeed satisfied. As the bands distort, and these distortions
are different for every choice of chemical potential $\mu_0$, it is clear that
the Fermi level will also need to be shifted. Again, to within the accuracy of
our numerical work, we have verified in a few instances that integration of the
occupied part of the states up to $\mu$ given by Eq.~(\ref{eq:interactmu})
does give the right charge imbalance $\rho$.

\begin{figure}[ht]
\begin{picture}(250,200)
\leavevmode\centering\includegraphics{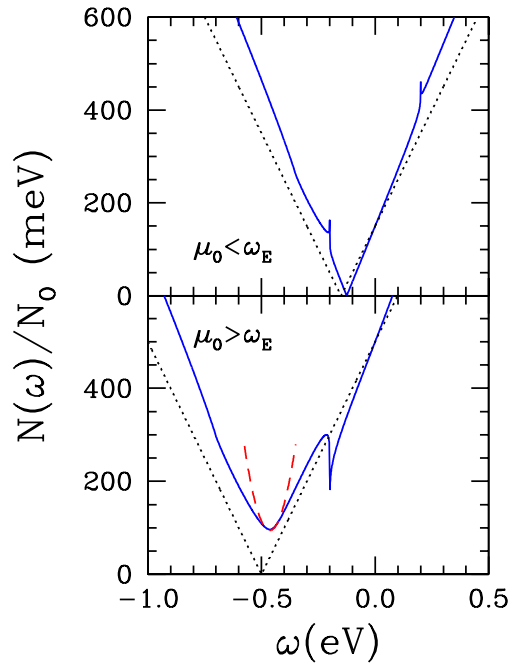}
\end{picture}
\vskip 50pt
\caption{(Color online)  The density of states $N(\omega)$ (solid line)
 vs. $\omega$ for $\omega_E=200$ meV, $\lambda=2.5$,
$W_C=7000$ meV. The top frame is for $\mu_0=150$ meV $<\omega_E$ and the
bottom for $\mu_0=500$ meV $>\omega_E$. The dotted curve is the bare band
case. For $\mu_0>\omega_E$, $N(\omega)$ at the Dirac point is nonzero and becomes
quadratic. A comparison of this approximate quadratic behavior 
given in Eq.~(\ref{eq:dosdirac}) is shown as the dashed (red) curve.
}
\label{Fig6}
\end{figure}

Returning to $\omega$ of order, at most, a few times $\omega_E$, a different
point is illustrated in Fig.~\ref{Fig6}, where we compare results for the 
DOS for the case of Fig.~\ref{Fig1} with $\mu_0<\omega_E$ (top frame)
with a case in which all parameters are kept the same but a value
of $\mu_0>\omega_E$ namely $\mu_0=500$meV is used. In the top frame
the DOS at the Dirac point remains zero because for $\mu_0=150$meV, the imaginary
part of $\Sigma(\omega)$ is still zero at the energy of the Dirac point as is
shown in the bottom frame of Fig.~\ref{Fig2}.
But in the bottom frame
 for $\mu_0=500$ meV, this is no longer the case and the Dirac point becomes
smeared and $N(\omega_d)$ is finite. Note that the shift between
Dirac point position for bare and interacting bands has increased as
compared to the case of the top frame (see middle frame of Fig.~\ref{Fig2}). 
It is useful to get a simple
analytic expression for the DOS around the Dirac point.

\begin{figure}[ht]
\begin{picture}(250,200)
\leavevmode\centering\includegraphics{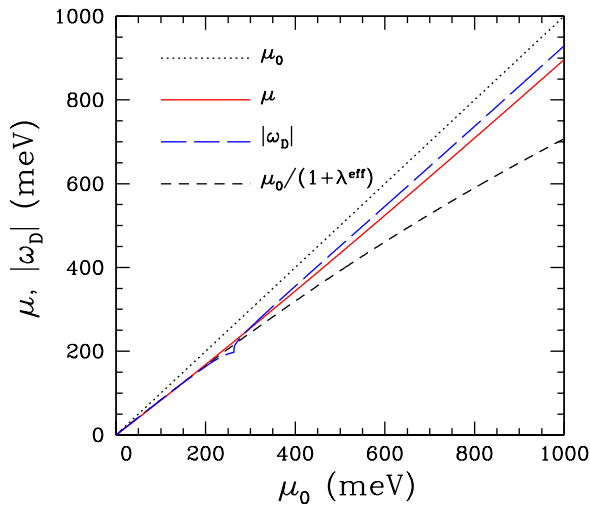}
\end{picture}
\vskip 0pt
\caption{(Color online)  Comparison of $\mu$ (red solid curve) and
$|\omega_d|$ (long-dashed blue curve) with approximate relation
$\mu_0/(1+\lambda^{\rm eff})$ (short-dashed black curve) as a function
of $\mu_0$. The dotted line is for comparison and gives $\mu_0$ unrenormalized.
}
\label{Fig7}
\end{figure}

For $\omega$ near $\omega_d$ and $\epsilon$ near $\epsilon=0$,
the spectral functions that determine the DOS are 
\begin{widetext}
\begin{equation}
A(\pm\epsilon,\omega)\simeq\frac{1}{\pi}\frac{-{\rm Im}\Sigma(\omega_d)}
{[\omega-{\rm Re}\Sigma(\omega_d)-{\rm Re}\Sigma^\prime(\omega_d)
(\omega-\omega_d)+\mu\pm\epsilon]^2+[{\rm Im}\Sigma(\omega_d)]^2}.
\label{eq:A}
\end{equation}
\end{widetext}
Denoting $1-{\rm Re}\Sigma^\prime(\omega_d)$ by $Z$, where 
$\Sigma^\prime(\omega_d)\equiv d\Sigma(\omega)/d\omega|_{\omega=\omega_d}$,
and $-{\rm Im}\Sigma(\omega_d)$ by $\Gamma$, we obtain
\begin{equation}
A(\pm\epsilon,\omega)\simeq\frac{1}{\pi}\frac{\Gamma}
{[(\omega-\omega_d)Z\pm\epsilon]^2+\Gamma^2}
\label{eq:Ared}
\end{equation}
and approximately (for $|(\omega-\omega_d)Z|\ll\Gamma$),\cite{sgrph}
\begin{equation}
\frac{N(\omega)}{N_0}=\frac{2\Gamma}{\pi}\ln\biggl|\frac{W_C}{\Gamma}\biggr|+
\frac{(\omega-\omega_d)^2Z^2}{\pi\Gamma},
\label{eq:dosdirac}
\end{equation}
where the first term is the value right at $\omega=\omega_d$
and the $\omega$ 
variation off the Dirac point is quadratic in $\omega-\omega_d$. 
Note that $N(\omega_d)
\to 0$ as $\Gamma\to 0$ in Eq.~(\ref{eq:dosdirac}). Also from Eq.~(\ref{eq:Ared}),
as $\Gamma\to 0$, $A(\pm\epsilon,\omega)\sim\delta[(\omega-\omega_d)Z\pm\epsilon]$
which gives for $\mu_0<\omega_E$ 
a linear variation of $N(\omega)$  out of $\omega_d$ with slope modified by
$Z$. We have tested  these analytic results against
numerical work and offer a comparison in 
the bottom frame of 
Fig.~{\ref{Fig6}}. The dashed red curve fits very well the solid blue curve
in the region of the Dirac point. To end this section, we comment on the range
of validity of the approximate but very useful Eq.~(\ref{eq:21}) and (\ref{eq:26})
for renormalized chemical potential and Dirac point position, respectively.
In Fig.~\ref{Fig7}, we compare exact results for $\mu$ (solid red curve)
and $\omega_d$ (long-dashed blue curve) with the approximation (\ref{eq:21})
and (\ref{eq:26}), respectively (i.e., $\mu_0/(1+\lambda^{\rm eff})$, short-black
curve). At small $\mu_0$, the agreement is excellent as we expected. The
deviations at higher values of $\mu_0\gtrsim 350$ meV
should be noted and could be of importance  in some applications.
Finally note that on the scale shown, $\mu$ and $|\omega_d|$ track each other
well.

\section{More realistic phonon spectra}

In a metal with constant density of states in the important energy
range for phonons, the electron-phonon interaction does not
renormalize its value.\cite{sgrph} The essential argument can be seen from Eq.~(\ref{eq:spectral}) 
which is to be integrated over energy, but $\int^{+\infty}_{-\infty}N(0)d\epsilon A(\epsilon,\omega)=N(0)$ independent of $\omega$.
This means that phonons will not show up in normal density of states
spectroscopy. But in graphene, this no longer holds for two reasons. First,
the Dirac density of states is linear in $\epsilon$ rather than constant
as can be seen in Eq.~(\ref{eq:dos}) and after integration $\Sigma$
remains in formula (\ref{eq:dosformula}). Second, we have finite bands and it has
been noticed by Knigavko et al.\cite{knigavko} that this also provides a mechanism
whereby an image of the phonon structure appears in the electronic density of
states. To examine phonon structure it is convenient to move away from
coupling to a single Einstein phonon as we have done so far in Eq.~(\ref{eq:Sigma})
for the self-energy. For a distribution of phonon energies $P(\nu)$,
it is necessary to average Eq.~(\ref{eq:Sigma}) over the desired distribution.
As a model, we use the truncated Lorentzian form of Ref.~\cite{frank}:
\begin{equation}
P(\nu)=
\frac{A'}{\pi}\biggl(\frac{\delta}{(\nu-\omega_0)^2+\delta^2}-\frac{\delta}{\delta_c^2+\delta^2}\biggr)\theta(\delta_c-|\omega_0-\nu|),
\label{eq:Pnu}
\end{equation}
where it is peaked at $\omega_0$ with width $\delta$ and truncated
at energy $\omega_0\pm\delta_c$. $A'$ is adjusted to give 
$\lambda$ used previously.
With this distribution
\begin{equation}
\Sigma_{\rm lor}(\omega)=\int_{-\infty}^{\infty} P(\nu)\Sigma(\omega,\nu)d\nu,
\label{eq:Sigmalor}
\end{equation}
where $\Sigma(\omega,\nu)\equiv\Sigma(\omega)$ of Eq.~(\ref{eq:Sigma})
and $\omega_E$ of that equation now becomes a variable $\nu$. Also,
the $A$ of Eq.~(\ref{eq:Sigma}) is omitted in favor of
Eq.~(\ref{eq:Pnu})
which enters into Eq.~(\ref{eq:Sigmalor}).
$\Sigma_{\rm lor}(\omega)$ is then used for the self-energy in the DOS
calculation 
using Eq.~(\ref{eq:dosformula}). In our calculations, we have used 
$\omega_0=200$ meV, $\delta=15$ meV, $\delta_c=30$ meV.

\begin{figure}[ht]
\begin{picture}(250,200)
\leavevmode\centering\includegraphics{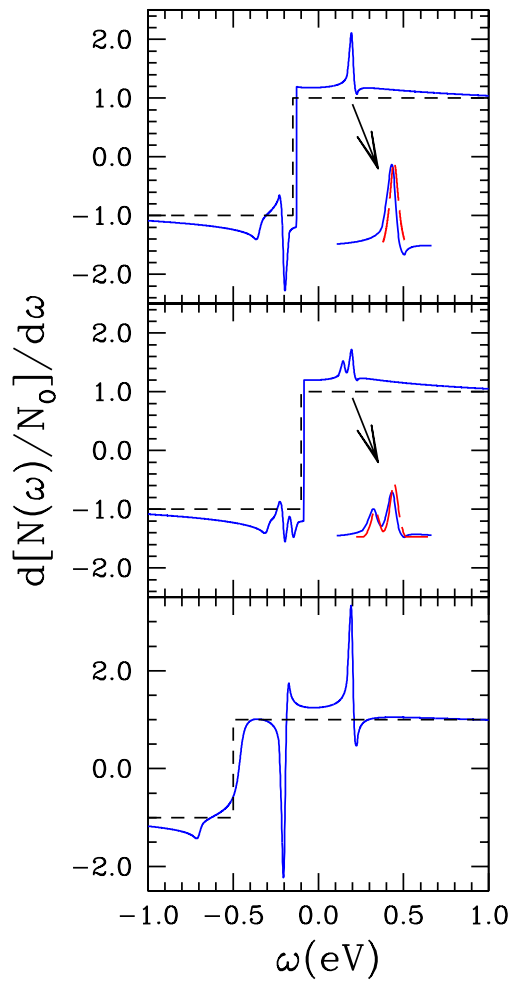}
\end{picture}
\vskip 160pt
\caption{(Color online) Derivative of the electronic density of states.
Dashed black line is the bare case and the solid blue curve includes phonons.
The top frame is for a bare chemical potential $\mu_0=150$ meV and coupling
to a Lorentzian phonon distribution peaked
at $\omega_0=200$ meV, while the middle is the
same but a distribution with two Lorentzian peaks
and $\mu_0=100$ meV. The insets indicated by an arrow, show the
structure
of the solid blue
curve
in comparison with the input electron-phonon spectral function (red
dashed curve). See text for further discussion.
The lower frame is also for the Lorentzian distribution used for the upper frame
 but now
$\mu_0=500$ meV.
}
\label{Fig8}
\end{figure}

To see better the phonon structures encoded in the electronic density of
states $N(\omega)$ of Figs.~\ref{Fig1} and \ref{Fig3}, it is convenient
to take a first derivative. This is shown in Fig.~\ref{Fig8} for several
cases. $d[N(\omega)/N_\circ]/d\omega$ is normalized such 
that in the bare band,
it is one at energies above the Dirac point and minus one below, with
a jump at  that point. This is represented
in Fig.~\ref{Fig8} as the black dashed curves with $\mu_0=150$ meV
for the top and middle frames and $\mu_0=500$ meV in the bottom frame.
The solid blue line contains the phonons. 
In the top frame, we see a prominent phonon structure at $\omega=\pm\omega_0$
in $d[N(\omega)/N_\circ]/d\omega$ which is 
superimposed on the bare band background
above which it rises by more than a factor of 2. 
On the negative energy side there is a further
smaller structure at $(-\mu_0-\omega_0)$. 
All three phonon structures can be traced to logarithmic singularities
in the self-energy of Eq.~(\ref{eq:Sigmamugt}) of the form $\ln|\omega\pm\omega_0|$ for the first two and $(\omega+\mu_0+\omega_0)\ln|\omega+\mu_0+\omega_0|$
for the last which is a weaker singularity.
Note that the vertical
drop in the solid curve is not at $\omega=-\mu_0$ but is rather
shifted to the right because the Dirac point has been shifted and this
drop signals the position of $\omega_d$ rather than of $-\mu_0$.  The
value of this derivative at $\omega=0$ has height equal to
$(1+\lambda^{\rm eff})$ and this can be used to measure this
renormalization parameter. Of course this is only possible if one knows
the value of the bare band Fermi velocity $v_0$.

On the
scale of this figure, it is clear that  the phonon structure should be easily
detectable. In the inset indicated by the solid arrow, we compare the structure
in the first derivative of the DOS (blue solid curve) with the input
spectral density (red dashed curve) scaled down by a factor of 10. Except
for a small shift between the two curves, there is excellent agreement and
conclude that the first derivative not only indicates where the boson
structure lies but also captures well the correct profile of the 
Lorentzian spectra that we have used. This is further emphasized in the
middle frame where two Lorentzians at slightly different energies
$\omega_{01}=150$ meV and $\omega_{02}=200$ meV are used with the second peak
height chosen to be twice that of the first. Again the reproduction of the details of the
input electron-phonon spectral density, which is scaled down by a factor of 10,
 are excellent.
This demonstrates
that normal density of states spectroscopy can potentially be used to
probe the phonon structures in graphene. The technique could also 
be extended to other structures
such as those due to 
hole-particle excitations and plasmons. The lower frame is the same
as the top  frame but now $\mu_0=500$ meV
beyond the phonon energies of $\sim 200$ meV in our model. In this case the
phonon structures are even bigger as we expect since the electron-phonon
interaction reflects the bare density of states which increases with
increasing energy. This expected increase in coupling can clearly be probed
in density of states measurements. A feature to be noted is that because
$\mu_0>\omega_0$ in this frame, the Dirac point is changed to the quadratic
behavior shown in the lower frame of Fig.~\ref{Fig6} and this leads to a drop in
$dN(\omega)/d\omega$ which is no longer vertical but instead 
is smeared. Finally, we
note that while the first derivative already gives a rather good image
of the shape of the underlying electron-phonon spectral function, a formal
inversion of Eq.~(\ref{eq:dosformula}) along the lines used in superconductivity
by McMillan and Rowell\cite{mcmillan} could be used to get its size and shape for each doping value.

\begin{figure}[ht]
\begin{picture}(250,200)
\leavevmode\centering\includegraphics{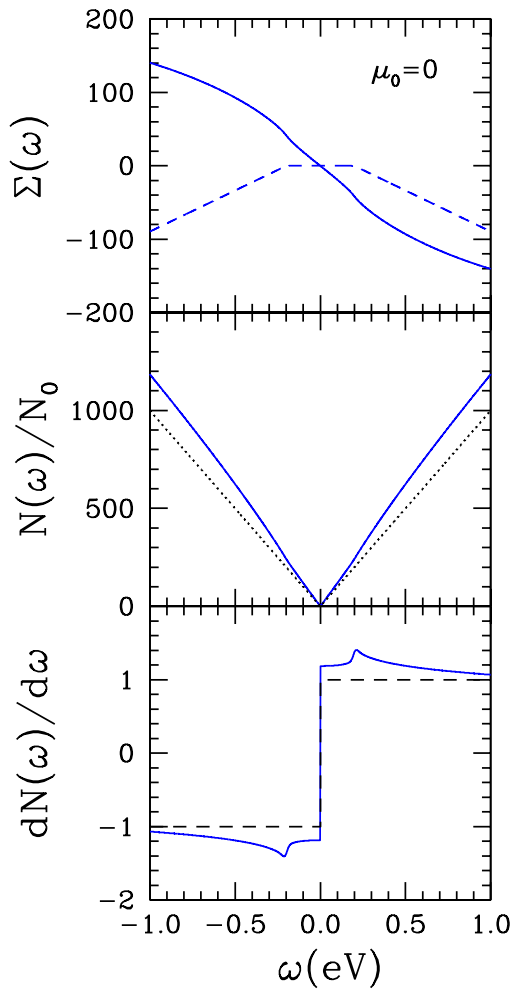}
\end{picture}
\vskip 160pt
\caption{(Color online) Top frame: Self-energy $\Sigma(\omega)$ for
  $\mu_0=0$
 in
units of meV for the real (solid curve) and imaginary part (dashed
curve). The density of states for this case is shown in the middle
frame in units of meV
with the bare DOS given as the dotted curve. 
Derivative of the DOS normalized to $N_0$ is shown in the bottom
frame with the
dashed black line is the bare case. In all cases, 
the solid blue curves includes phonons.
}
\label{Fig9}
\end{figure}

As we noted with reference to the top frame of Fig.~\ref{Fig1}, the logarithmic
singularities at $\omega=\pm\omega_E$ in Eq.~(\ref{eq:Sigmamugt}) for
${\rm Re}\Sigma(\omega)$ no longer appear when the chemical potential
$\mu_0=0$ and the bands are filled exactly to the Dirac point. This is
shown in Fig.~\ref{Fig9} (top frame). Only a small ``kink''
(i.e., change in slope)
remains at $\omega=\pm\omega_E$. This fact
has important consequences for experiment. While the density of states
as a function of $\omega$ still knows about the renormalization factor of
$1+\lambda^{\rm eff}$ (middle frame) 
 there is no additional logarithmic-type signature
of the Einstein phonon mode. Consequently, compared with the
examples of Fig.~\ref{Fig8}, only very small
peaks are seen at $\omega=\pm\omega_E$ in the derivative, as shown 
in Fig.~\ref{Fig9} (bottom frame). Nevertheless, one can see the
$1+\lambda^{\rm eff}$ renormalization at low frequency in this frame,
given
as the magnitude of the height of the curve near zero frequency.
In their scanning tunneling microscope (STM)
results, Li et al.\cite{li} find structure in the DOS at 155 meV. 
In addition, they find a renormalization which they determine to be
$\lambda^{\rm eff}=0.26$, for a case of doping very close to
the charge neutrality point. This case was addressed in our previous short
communication\cite{sgrph} which was aimed towards a detailed
evaluation of the data analysis procedure used by Li et al.
for extracting a value of the mass enhancement. From that we conclude that
the data may support a larger value of $\lambda$, possibly as large as
0.4.

\section{Angle-Resolved Photoemission (ARPES)}

\begin{figure}[ht]
\begin{picture}(250,200)
\leavevmode\centering\includegraphics{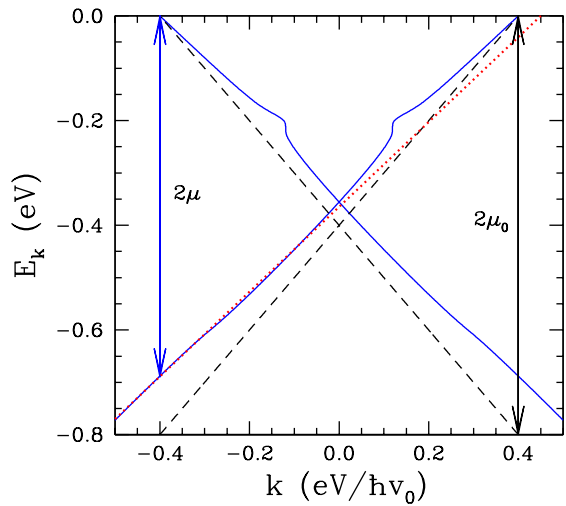}
\end{picture}
\vskip 20pt
\caption{(Color online) Renormalized energies (solid blue curves)
$E_k$ as a function of $k$ in units of eV$/(\hbar v_0)$ for a case with $\mu_0=400$ meV.
Twice the bare and dressed chemical potential are indicated by vertical
arrows, long and short, respectively. The
bare curves are shown as dashed black lines and the dotted is a line
chosen
to fit the dressed curve asymptotically at large negative energy.
}
\label{Fig10}
\end{figure}

The charge carrier spectral density $A({\bf k},\omega)$ given
by Eq.~(\ref{eq:spectral}) can be measured in angle-resolved
photoemission
experiments (ARPES). The technique is directional and measures
$A({\bf k},\omega)$  as a function of $\omega$ for any direction
and magnitude of momentum ${\bf k}$. Here, because of the symmetry
only $\epsilon=\pm\hbar v_0|{\bf k}|$ enters. As we have already noted, if the
imaginary part of the electron-phonon self-energy is infinitesimal,
$A(\pm\epsilon,\omega)$  would reduce to a delta function and provide the 
dressed energy $E_k$ corresponding to the bare $\epsilon_k$ through
the equation $E_k-{\rm  Re}\Sigma(E_k)+\mu-\epsilon_k=0$. In
Fig.~\ref{Fig10}, we show results for $E_k$ in the case of
$\mu_0=400$
meV. The dashed black lines are the bare dispersions and the solid
blue curves, the renormalized energies. 
Several features of this figure
need to be noted. The renormalized dispersions show a clear ``kink''
at the peak phonon energy $\omega_0=200$ meV. In the calculations, we used
our Lorentzian model for the distribution of phonon energies about
$\omega_0$
otherwise the ``kink'' would actually show a logarithmic singularity.
For a simple metal with a constant DOS around the Fermi energy, the
same phonon anomaly would arise, but at higher energies (below the
Fermi energy) the renormalized curve would rapidly return to the bare
value as the real part of $\Sigma(\omega)$ drops to zero. This is not
true for graphene because the density of electron states increases
linearly with increasing energy and this means that 
$|{\rm  Re}\Sigma(\omega)|$ increases as we saw in the top frame of
 Fig.~\ref{Fig1}. The
phonon
structures are superimposed on this more gradual increase in 
$|{\rm  Re}\Sigma(\omega)|$ versus $\omega$. The dotted red line in
Fig.~\ref{Fig10}
brings out this feature clearly. It has been drawn to match the
dressed dispersions at large negative energies. In contrast to what
would be the case for the bare dispersions, this asymptotic
line when extended towards the Fermi energy does not cross through the
Fermi energy. This has been noted in the ARPES data of Bostwick et al.\cite{bostwick}.
In our calculations this is a direct consequence of the fact that the
electron-phonon renormalizations continue to increase at high energies
because of the continuing increase of the bare density of states.
In the actual experimental data,
other renormalizations would also contribute. While the slope of the
bare dispersion curves (dashed black lines) gives directly the bare
velocity
$v_0$ the slope of the asymptotic dotted line is smaller and
would
give a smaller value of the effective
velocity of light for graphene should it be used to define the bare bands.
It is close but not the same as the renormalized effective velocity of light
$v_0/(1+\lambda^{\rm eff})$.
 The vertical arrows
in
Fig.~\ref{Fig10} identify renormalized (shorter blue arrow) and bare
(longer
black arrow) chemical potentials, respectively. It is clear that $\mu$
is
about 15\%  smaller. This is close but does not quite correspond to
the
factor of $1+\lambda^{\rm eff}=1.19$ in our calculations indicated in
Eq.~(\ref{eq:21}) which is strictly valid only for $\mu_0\ll\omega_0$.
Nevertheless, the differences are small. Here the factor of 1.19
gives exactly the slope of $E_k$ out of the Fermi energy.

\begin{figure}[ht]
\begin{picture}(250,200)
\leavevmode\centering\includegraphics{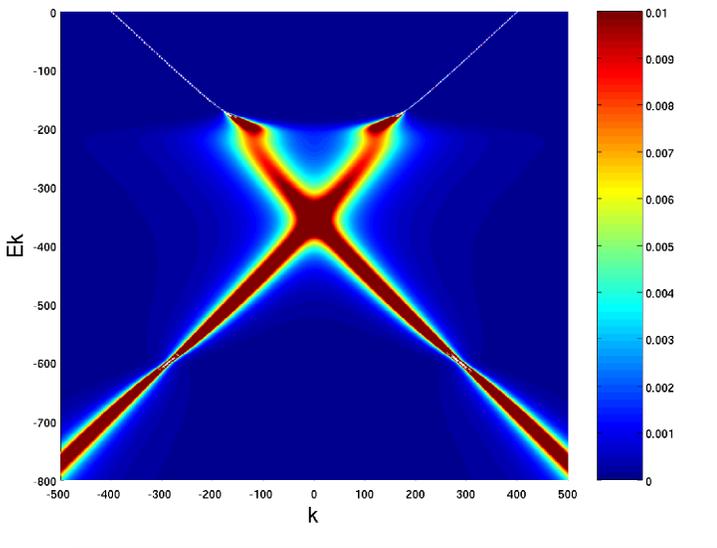}
\end{picture}
\vskip 20pt
\caption{(Color online) Color map of interacting dispersions $E_k$ in units
of meV 
as a function of ${\bf k}$ in units of  meV$/(\hbar v_0)$. The bare chemical potential
$\mu_0=400$ meV and a Lorentzian phonon spectrum with $\omega_0=200$
meV
were used.
}
\label{Fig11}
\end{figure}

Of course, the imaginary part of $\Sigma(\omega)$ in
Eq.~(\ref{eq:spectral})
for $\omega\ne 0$ would never be zero and we would not have perfectly
well-defined dressed dispersion curves. In such a circumstance, 
it is useful to
use
a color map to show the constant energy contours of
$A(\epsilon,\omega)$. This is done in Fig.~\ref{Fig11}. Cuts at
different
values correspond to different colors as indicated on the scale
shown
on the right (units of meV$^{-1}$). 
Because we have used a phonon spectrum which starts at 
$\omega=170$ meV for this figure ($\omega_0-\delta_c$), 
there is no quasiparticle lifetime
above $-170$ meV. Below this energy the curves widen and reflect the
broadening. We see clearly the peak of the phonon anomaly at $-200$ meV
but the identification of the position of the 
Dirac point becomes somewhat ambiguous because of the broadening of
the curves. The renormalized
$|\mu|=344$ meV for this case and according to
Fig.~\ref{Fig7} this almost coincides with
$|\omega_d|$. The pinching at $-600$ meV is due to the small
value of the imaginary part of the self-energy
 at $-\mu_0-\omega_0$ as we have already seen in the
middle
frame of Fig.~\ref{Fig1} for a different value of chemical
potential. Other
interactions such as electron-hole particle pair formation or coupling
due
to plasmons are expected to lift this pinching and provide a finite
lifetime. It is clear from this figure that the
constructions used in Fig.~\ref{Fig10} to identify the dotted
asymptotic
line as well as the interacting chemical potential are not as
well-defined when broadening is included in the theory.

\section{Electron-Phonon Interaction on DC Conductivity}

Next we turn to the optical conductivity.
In units of the universal background conductivity $\sigma_0=\pi e^2/2h$,
where $e$ is the electron charge and $h$ is Planck's constant, the real part of
the frequency-dependent conductivity $\sigma(T,\Omega)$ at frequency $\Omega$ and
temperature $T$ is given by
\begin{widetext}
\begin{equation}
\frac{\sigma(T,\Omega)}{\sigma_0}=
\frac{4}{\Omega}\int^{\infty}_{-\infty}d\omega
[f(\omega)-f(\omega+\Omega)]
\int_0^{W_C}\epsilon d\epsilon[A(\epsilon,\omega)+A(-\epsilon,\omega)]
[A(\epsilon,\omega+\Omega)+A(-\epsilon,\omega+\Omega)],
\label{eq:cond}
\end{equation}
\end{widetext}
where we have ignored vertex corrections. In metal physics, this can 
be incorporated approximately by an extra weighting of $(1-\cos\theta)$
in the calculation of the scattering rate changing it from a quasiparticle
to transport rate.\cite{Grimvall} This has recently been verified to hold as well for the specific
case of the DC conductivity of 
graphene by  Cappelluti and Benfatto\cite{vertexcorr}.
 The function
$f(\omega)=1/[\exp(\beta\omega)+1]$ is the Fermi occupation factor
with $\beta$, the inverse temperature. 
We start with a general observation that 
\begin{equation}
\sigma_{\mu<0}(T,\Omega)=\sigma_{|\mu|}(T,\Omega).
\label{eq:sigmamuneg}
\end{equation}
To establish this symmetry, we note first that for negative values of
the chemical potential $\mu$, the carrier spectral function satisfies
\begin{equation}
A_{\mu<0}(\epsilon,\omega)=A_{|\mu|}(-\epsilon,-\omega),
\label{eq:Amuneg}
\end{equation}
where we have used Eqs.~(\ref{eq:resigma}) and (\ref{eq:imsigma}).
Also, Eq.~(\ref{eq:cond}) can be rewritten as
\begin{widetext}
\begin{equation}
\frac{\sigma_{\mu<0}(T,\Omega)}{\sigma_0}=\displaystyle
\frac{4}{\Omega}\int^{\infty}_{-\infty}d\omega
[f(\omega)-f(\omega+\Omega)]
\int_0^{W_C}\epsilon d\epsilon[A(\epsilon,-\omega)+A(-\epsilon,-\omega)]
[A(\epsilon,-\omega-\Omega)+A(-\epsilon,-\omega-\Omega)].
\label{eq:condmuneg}
\end{equation}
\end{widetext}
Noting Eq.~(\ref{eq:Amuneg}), then Eq.~(\ref{eq:sigmamuneg}) follows.

Taking $\Omega=0$ in Eq.~(\ref{eq:cond}) gives the DC conductivity
at finite $T$,
\begin{widetext}
\begin{equation}
\frac{\sigma_{DC}(T)}{\sigma_0}=
-4\int^{\infty}_{-\infty}d\omega
\frac{\partial f(\omega)}{\partial \omega}
\int_0^{W_C}\epsilon d\epsilon[A^2(\epsilon,\omega)+A^2(-\epsilon,\omega)
+2A(\epsilon,\omega)A(-\epsilon,\omega)].
\label{eq:condDC}
\end{equation}
\end{widetext}
We have verified numerically that in Eq.~(\ref{eq:condDC}), we can
send $W_C$ to infinity within good approximation. This simplifies
the result and we get
\begin{widetext}
\begin{equation}
\frac{\sigma_{DC}(T)}{\sigma_0}=
-\frac{4}{\pi^2}\int^{\infty}_{-\infty}d\omega
\frac{\partial f(\omega)}{\partial \omega}
\biggl[1+\biggl(\frac{a(\omega)}{\Gamma(\omega)}+\frac{\Gamma(\omega)}{a(\omega)}
\biggr)\tan^{-1}\biggl(\frac{a(\omega)}{\Gamma(\omega)}\biggr)\biggr]
\label{eq:condDCsimple}
\end{equation}
\end{widetext}
with $\Gamma(\omega)=-{\rm Im}\Sigma(\omega)$ and $a(\omega)=\omega-{\rm Re}\Sigma(\omega)+\mu$.
Note that for $\Sigma(\omega)$, we have neglected the temperature
dependence that would enter Eq.~(\ref{eq:Sigma}) at finite
temperature\cite{frank}
and hence our results here are for low $T$.
At zero temperature $-\partial f(\omega)/\partial\omega$ in Eq.~(\ref{eq:condDCsimple})
becomes a Dirac delta function centered around $\omega=0$ (i.e.,
$\delta(\omega)$)
and 
\begin{widetext}
\begin{equation}
\frac{\sigma_{DC}(T=0)}{\sigma_0}=
\frac{4}{\pi^2}
\biggl[1+\biggl(\frac{\mu-{\rm Re}\Sigma(\omega=0)}{\Gamma(\omega=0)}+\frac{\Gamma(\omega=0)}{\mu-{\rm Re}\Sigma(\omega=0)}
\biggr)\tan^{-1}\biggl(\frac{\mu-{\rm Re}\Sigma(\omega=0)}{\Gamma(\omega=0)}\biggr)\biggr].
\label{eq:condDCT0}
\end{equation}
\end{widetext}
But $\mu-{\rm Re}\Sigma(\omega=0)$ in (\ref{eq:condDCT0}) can be replaced
by $\mu_0$, the bare band chemical potential from Eq.~(\ref{eq:interactmu}).
Also, at $\omega=0$ and $T=0$, the electron-phonon interaction does not
contribute to the scattering rate which therefore reduces to the
residual scattering rate $\eta$, i.e., $\Gamma(\omega=0)\equiv\eta$, which is small but finite for graphene. Thus,
the electron-phonon interaction has entirely dropped out of Eq.~(\ref{eq:condDCT0}), which reduces to its bare band value. This means that the
familiar result\cite{Grimvall,Prange} that the electron-phonon interaction
does not change the DC conductivity of ordinary metals also holds for 
graphene.

For $\mu-{\rm Re}\Sigma(\omega=0)\equiv\mu_0\ll\eta$ in Eq.~(\ref{eq:condDCT0}),
we get the universal DC limit,\cite{shon,gorbar} unrenormalized by the electron-phonon
interaction
\begin{equation}
\sigma_{DC}(T=0)=
\frac{8}{\pi^2}\sigma_0=\frac{4e^2}{\pi h}
\label{eq:condDCuniv}
\end{equation}
which differs from the universal AC background value $\sigma_0$ by a factor
of $8/\pi^2\simeq 0.81$.
The first correction for finite $\mu_0$ is included when Eq.~(\ref{eq:condDCuniv}) is multiplied by $(1+(\mu_0/\eta)^2/2)$.
 In the opposite limit of
$\mu-{\rm Re}\Sigma(\omega=0)\gg\eta$, we get instead
\begin{equation}
\sigma_{DC}(T=0)=\frac{e^2}{h}
\frac{2[\mu-{\rm Re}\Sigma(\omega=0)]}{2\eta}=
\frac{e^2}{h}\frac{2\mu_0}{2\eta}.
\label{eq:condDCother}
\end{equation}
Here,
$2\mu_0$ is the optical spectral weight
removed from the universal background at finite chemical potential
(in the bare band)
which reappears as a 
Drude type contribution about $\Omega=0$. Note that
 $1/\tau=2\eta$ is the transport 
scattering rate. In this limit, the DC conductivity is no longer universal,
i.e., $\eta$ does not drop out as it did in Eq.~(\ref{eq:condDCuniv}),
 however the electron-phonon interaction has dropped out.
This is similar to the well-known result\cite{Grimvall,Prange} in conventional metals, that the
electron-phonon interaction does not change the value of the DC conductivity. 
This  remains
true for graphene at $T=0$, and can be traced to 
Eq.~(\ref{eq:condDCT0}).
 $\sigma_{DC}(T=0)$ 
 does not depend on the electron-phonon interaction
because of 
the appearance of the chemical potential 
shift factor ${\rm Re}\Sigma(\omega=0)$,
which changes $\mu$ to $\mu_0$, the bare band value, and $\Gamma(\omega=0)$
reduces to the residual scattering rate.

\begin{figure}[ht]
\begin{picture}(250,200)
\leavevmode\centering\includegraphics{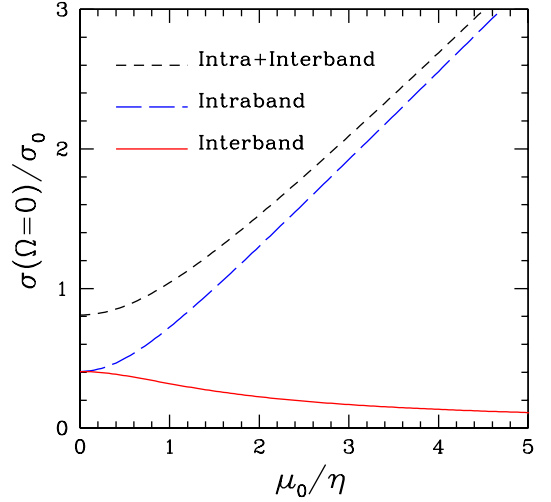}
\end{picture}
\vskip 20pt
\caption{(Color online) DC conductivity $\sigma(\Omega=0)$ at $T=0$
in units of the universal background value $\sigma_0=\pi e^2/2h$. The
solid red curve gives the interband contribution, the long-dashed blue curve,
 the intraband and the short-dashed black curve, the sum as a function of
$[\mu-{\rm Re}\Sigma(\omega=0)]/\eta\equiv\mu_0/\eta$, with $\mu$ the chemical
potential, $\mu_0$ its bare band value,
${\rm Re}\Sigma(\omega=0)$ its shift, and
 $\eta$ the residual quasiparticle scattering rate.
}
\label{Fig12}
\end{figure}

Returning to 
Eq.~(\ref{eq:condDC}), it is useful to separate
intraband and interband contributions. The first two terms are intraband
because they involve transitions within the same Dirac cone while the last
term involves transitions between lower and upper cones. 
In a conventional metal only the intraband
contribution arises in discussions of the DC conductivity. In Fig.~{\ref{Fig12}},
we show results for $\sigma_{DC}(T=0)$ in units of $\sigma_0$ as a function 
of $(\mu-{\rm Re}\Sigma(\omega=0))/\eta\equiv\mu_0/\eta$. 
The solid (red) curve is the interband piece,
the long-dashed (blue) curve the intraband and the short-dashed (black), the sum.
Both the intraband and interband transitions contribute equally to
the universal DC value which corresponds to $(\mu-{\rm Re}\Sigma(\omega=0))/\eta=\mu_0/\eta=0$,
i.e., the lower cone is filled to the Dirac point and the upper cone is empty. 
As $\mu_0$ is increased and the material is doped away
from the neutrality point, the interband  contribution
decreases while at the same time the sum increases rather
substantially. What determines the scale of this increase is the value
of the ratio of the
chemical potential to the residual scattering rate $\eta$. While the
value of $\eta$ is somewhat uncertain for graphene and will vary with sample
quality, it is expected to be small and of order of 1.0 meV so that the value
of the DC conductivity can be increased a lot for easily attained values of the
chemical potential, say $\mu_0$ of order a few hundred meV. 

Returning to Eq.(\ref{eq:condDCsimple}) which gives the DC conductivity
at finite temperature $T$, we note that the value of ${\rm Re}\Sigma(\omega)$
for finite $\omega$ now enters the formula even for $\mu=0$ (Dirac point is at the 
Fermi surface) and consequently the DC conductivity is affected by the electron-phonon
interaction. It is only for $T=0$, $\mu_0=0$ that we get $\sigma_{DC}=0.81\sigma_0$ independent of electron-phonon
renormalization.

\begin{figure}[ht]
\begin{picture}(250,200)
\leavevmode\centering\includegraphics{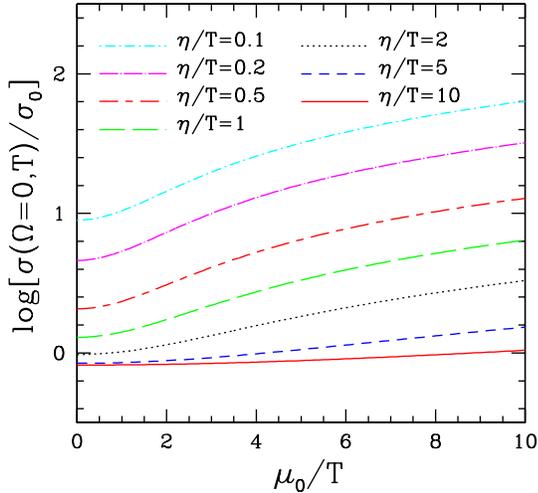}
\end{picture}
\vskip 20pt
\caption{(Color online) The DC conductivity $\sigma(\Omega=0,T)$  
at finite, but small, temperature $T$ in units of $\sigma_0=\pi e^2/2h$ as a function of
$\mu_0/T$ for several values of $\eta/T$ with $\mu_0$, the bare chemical
potential and $\eta$, the residual quasiparticle scattering rate.
Here, for simplicity, phonon renormalizations are taken to be zero.
}
\label{Fig13}
\end{figure}

In a conventional metal, if only impurity (elastic) scattering is accounted
for, the DC conductivity does not change with increasing temperature. To
get a change it is necessary to include inelastic processes.
 This is not so for graphene. Including only
residual scattering and no electron-phonon renormalization, the formula
for the DC conductivity  Eq.~(\ref{eq:condDCsimple}) reduces to 
\begin{widetext}
\begin{equation}
\frac{\sigma_{DC}(T)}{\sigma_0}=
\frac{4}{\pi^2}\int^{\infty}_{-\infty}\frac{dx}{4\cosh^2[(x-\tilde\mu)/2]}
\biggl[1+\biggl(\frac{x}{\tilde\eta}+\frac{\tilde\eta}{x}
\biggr)\tan^{-1}\biggl(\frac{x}{\tilde\eta}\biggr)\biggr].
\label{eq:condDCnoeph}
\end{equation}
\end{widetext}
In Fig.~\ref{Fig13}, we show on a logarithmic scale $\sigma(\Omega=0,T)/\sigma_0$
as a function of $\tilde\mu\equiv\mu/T$ for various values of 
$\tilde\eta\equiv\eta/T$. Here, $\mu=\mu_0$, the bare band chemical potential,
as $\Sigma(\omega)$ has been taken to be zero. We first note that at
$\tilde\mu=0$, the universal limit is no longer universal unless the temperature
is much smaller than the residual quasiparticle scattering rate $\eta$. 
An analytic expression valid for $T/\eta\ll 1$ is\cite{beck} 
\begin{equation}
\sigma_{DC}(T)=
\frac{4e^2}{\pi h}\biggl[1+\frac{\pi^2}{9}\biggl(\frac{T}{\eta}\biggr)^2\biggr].
\end{equation}
For
$\eta=T$, the DC conductivity has increased by more than a factor of two over
its $T=0$ value. It is clear that in clean systems, it is necessary to go to 
very low temperatures in order to observe the universal limit. It is also clear
that the chemical potential is to be small compared with $\eta$. It is the
residual scattering rate which sets the scale on $T$ and $\mu$. The conditions
to observe the universal limit are $T\ll\eta$ and $\mu\ll\eta$. We turn next to 
finite $\Omega$ in the microwave or Terahertz region where $\Omega$ is of order
$\eta$, and study the evolution of the conductivity from its DC value
to its universal AC background value.

\section{Low frequency conductivity}

\begin{figure}[ht]
\begin{picture}(250,200)
\leavevmode\centering\includegraphics{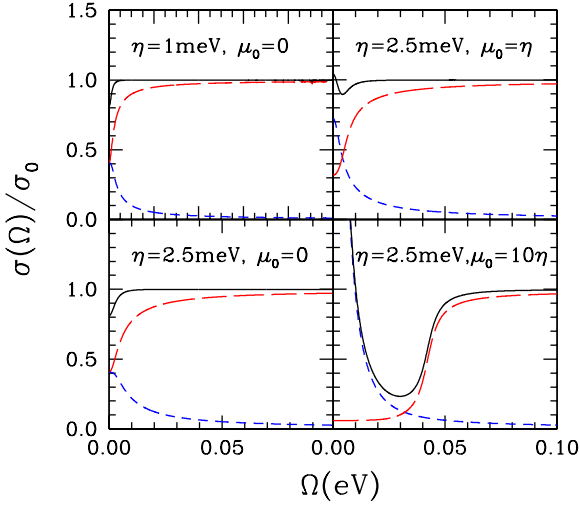}
\end{picture}
\vskip 20pt
\caption{(Color online) The conductivity $\sigma(\Omega)$ at 
$T=0$ in units of $\sigma_0=\pi e^2/2h$ as a function of
$\Omega$  emphasizing the low frequency region. Left frames are for zero
chemical potential $(\mu_0=0)$ and quasiparticle residual scattering
rate $\eta=1.0$ and 2.5 meV. Right frames are for $\eta=2.5$ meV with
$\mu_0=\eta$ and $10\eta$. Long-dashed red curve is the interband
contribution, short-dashed blue is the intraband and solid black is
the sum. Electron-phonon renormalizations are included using the
parameters
previously discussed.
}
\label{Fig14}
\end{figure}

In Fig.~\ref{Fig14}, we show results for $\sigma(\Omega)$ at $T=0$ in the
frequency region spanning a few times the energy of the quasiparticle
scattering rate $\eta$. The top left hand frame shows $\sigma(\Omega)/\sigma_0$
versus $\Omega$ up to 100 meV for zero chemical potential with $\eta=1$ meV
while the lower frame is the same but now $\eta=2.5$ meV. In both cases
the long dashed (red) curve is the interband contribution and the short-dashed 
(blue) is the intraband. The sum is the solid black curve. We see that
at $\Omega=0$, the universal limit is reached for both
values of $\eta$ as expected. As $\Omega$ is increased $\sigma(\Omega)$ increases
toward the universal background value $\sigma_0=\pi e^2/(2h)$. The energy scale on which saturation is reached
is set by the value of $\eta$ and hence will vary with sample quality. The two right hand panels are similar
but 
are for finite chemical potential,
with $\mu_0=\eta$ for the top and $\mu_0=5\eta$ for the bottom, and
$\eta=2.5$ meV in both cases. It is clear that the conductivity in the low
$\Omega$ region is greatly affected by finite $\mu_0$ and is non-universal. 
This arises because at finite $\mu_0$, the interband transitions below $2\mu_0$
are Pauli blocked and the optical spectral weight involved is transferred to the
intraband transitions. This provides a Drude peak with finite effective
plasma frequency proportional to $2\mu_0$. In the top right panel of 
Fig.~\ref{Fig14}, $\mu_0=\eta$ so that we are not in the universal
DC limit regime and its value is in fact already larger than the universal
background value $\sigma_0$. As $\Omega$ is increased, $\sigma(\Omega)/\sigma_0$
drops, shows a minimum before rising again to reach its universal
background value.
The energy scale for this final rise to saturation remains $\eta$ as in the
left hand panels but this is because $\eta$ is of order $\mu_0$. For
a case $\mu_0>\eta$ as in the lower right panel, specifically with 
$\mu_0=10\eta$, the Drude at small $\Omega$ is much more pronounced as compared
with the upper frame and it is now $2\mu$ which sets the energy scale for
recovery of the conductivity to its universal background value. An interesting
feature of the interband contribution to the conductivity (long-dashed red
curve) in this case is the long tail extending to $\Omega=0$ which is
nearly constant before its main rise around $\Omega=2\mu$. This behavior
can be understood analytically as we now show. 

At zero temperature, for small but finite $\Omega$ the conductivity is given
approximately by
\begin{widetext}
\begin{equation}
\frac{\sigma(\Omega)}{\sigma_0}=\frac{4}{\Omega}\int_{-\Omega}^0d\omega\sum_{\pm}\int_0^\infty
\frac{\epsilon d\epsilon}{\pi^2}\biggl[\frac{\eta}{[\omega(1+\lambda_{\rm eff})
\pm\epsilon+\mu_0]^2+\eta^2}\biggr]\biggl[\frac{\eta}{[(\omega+\Omega)(1+\lambda_{\rm eff})\pm\epsilon+\mu_0]^2+\eta^2}\biggr]
\label{eq:intra}
\end{equation}
\end{widetext}
where we have noted the $\omega$ is also small and have used
$-\lambda_{\rm eff}\omega$ for
${\rm Re}\Sigma(\omega)$, after subtraction of the constant piece. 
The imaginary part of the self-energy due to phonons is zero so that only the
residual contribution remains. Should the acoustic phonons have been
included in the model, ${\rm Im}\Sigma(\omega)$ would be finite, but
this effect
is small
in graphene and neglected here. 
The sum over $\pm$ leads to four terms in Eq.~(\ref{eq:intra}),
two intraband contributions (same sign) and two interband (opposite
signs of $\epsilon$ in pair of terms).
The integral over energy can be performed analytically 
and the sum over $+$ and $-$, i.e., upper and lower
Dirac cone performed. Starting first with the intraband piece and writing
$\bar\Omega\equiv\Omega(1+\lambda^{\rm eff})$ and $\bar\omega\equiv\omega(1+\lambda^{\rm eff})$,
we get
\begin{widetext}
\begin{eqnarray}
\frac{\sigma_{intra}(\Omega)}{\sigma_0}&=&\frac{4\eta^2}{\pi^2\Omega}\int_{-\Omega}^0d\omega\biggl\{\frac{-2}{\bar\Omega}
\frac{1}{4\eta^2+\bar\Omega^2}\biggl[\{(\bar\omega+\frac{\bar\Omega}{2})+\mu_0\}\ln\bigg|\frac{[\bar\omega+\mu_0]^2+\eta^2}
{[(\bar\omega+\bar\Omega)+\mu_0]^2+\eta^2}\biggl|\biggl]\nonumber\\
&&-4\eta-\frac{\bar\Omega}{\eta}[(\bar\omega+\bar\Omega)+\mu_0]\tan^{-1}\biggl(\frac{(\bar\omega+\bar\Omega)}{\eta}+\mu_0\biggr)+\frac{\bar\Omega}{\eta}(\bar\omega+\mu_0)
\tan^{-1}\biggl(\frac{\bar\omega}{\eta}+\mu_0\biggr)\biggr\}.\nonumber\\
\label{eq:condlong}
\end{eqnarray}
\end{widetext}
Assuming $\omega,\Omega, \eta\ll\mu_0$, expression (\ref{eq:condlong}) greatly
reduces and we get
\begin{equation}
\frac{\sigma_{intra}(\Omega)}{\sigma_0}=\frac{4|\mu_0|}{\pi}\frac{2\eta}{\Omega^2(1+\lambda^{\rm eff})^2+4\eta^2}
\label{eq:condintra}
\end{equation}
for the intraband contribution to the conductivity when $\Omega\ll\mu_0$,
as well as, $\eta\ll\mu_0$.
The form (\ref{eq:condintra}) is a Drude with effective optical
scattering rate $2\eta/(1+\lambda^{\rm eff})$ and effective plasma frequency of
$4\mu_0/[\pi(1+\lambda^{\rm eff})]$. While we have made approximations
to derive Eq.~(\ref{eq:condintra}), we have verified in numerical work for Fig.~\ref{Fig14}
right bottom frame that the long-dashed blue curve follows (\ref{eq:condintra})
very well. The result (\ref{eq:condintra}) is the same as found in conventional
metals. The coherent part of the conductivity is a Drude form
with scattering rate and plasma
frequency renormalized by $(1+\lambda^{\rm eff})$ and the DC limit remains unrenormalized by
$\lambda^{\rm eff}$. Here  for simplicity, we have used a constant
residual
scattering rate. For graphene, this scattering rate can itself be
frequency
dependent and this can change the Drude line shape which then reflects
this additional energy dependence as described in Ref.~\cite{gusynin}.

\begin{figure}[ht]
\begin{picture}(250,200)
\leavevmode\centering\includegraphics{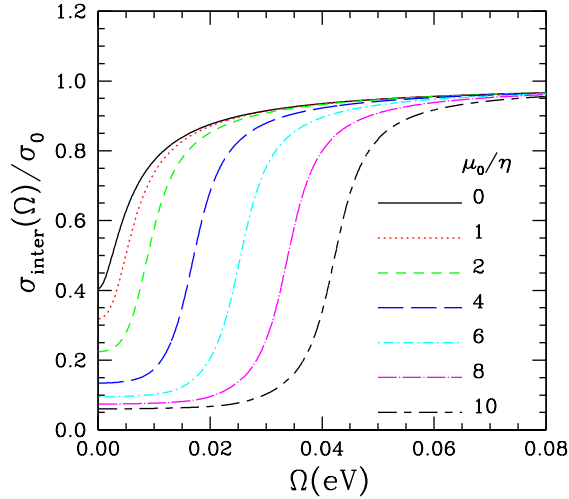}
\end{picture}
\vskip 30pt
\caption{(Color online) The interband conductivity $\sigma_{inter}(\Omega)$ at 
$T=0$ in units of $\sigma_0=\pi e^2/2h$ as a function of photon energy
$\Omega$  emphasizing the low frequency region, for varying values of
  $\mu_0/\eta$, with $\eta=2.5$ meV. Electron-phonon renormalizations
  are
included as described previously.
}
\label{Fig15}
\end{figure}

Application of the same sort of algebra for the interband contribution to the
conductivity in the limit $\Omega$ and $\eta\ll\mu_0$, gives a leading
contribution which is {\it frequency 
independent}:
\begin{equation}
\frac{\sigma_{inter}(\Omega)}{\sigma_0}
=\frac{1}{\pi}\frac{2\mu_0\eta}{\eta^2+\mu_0^2},
\label{eq:condinter}
\end{equation}
$(\Omega<\omega_E)$
and in sharp contrast to Eq.~(\ref{eq:condintra}) is independent of the
electron-phonon renormalization parameter $\lambda^{\rm eff}$. Thus,
while the intraband Drude contribution is renormalized by the electron-phonon
interaction in the same way as in ordinary metals, the interband is not. 
Equally remarkable is that (\ref{eq:condinter}), in contrast to (\ref{eq:condintra}), is independent of photon energy, 
as we already noted in the lower right hand 
frame of Fig.~\ref{Fig14} where we present numerical results.
The interband contribution (long-dashed red curve) remains essentially 
constant at its DC value even  for $\Omega\lesssim 2\mu$  at which point it
begins a steep rise and eventually makes the major contribution to the
universal
background value. At small $\Omega$ it is the reverse that holds, the
interband
piece is small while the intraband Drude-like contribution is
dominant.
Additional results for the interband contribution to the conductivity
for photon energies less than and slightly above the value of twice
the  chemical potential are shown in Fig.~\ref{Fig15}. Here we span
values of $\mu_0$ from 0 to 10 times $\eta$, as labeled on the
figure. Only the first curve (solid black) satisfies the condition
for the universal DC limit and in this case, the curve 
starts at a value of $4/\pi^2\sim 0.4$ and makes half the contribution
to the universal DC conductivity. Furthermore, it rapidly rises
to its universal AC background value on the scale of $2\eta$, with 
$\eta=2.5$ meV. All other curves start with
$\sigma(\Omega)/\sigma_0$ below $4/\pi^2$ at $\Omega=0$. 
This is true even for the red dotted curve for which $\mu_0=\eta$.
 The last few conform
well to the analytic prediction of Eq.~(\ref{eq:condinter})
that $\sigma_{inter}(\Omega)/\sigma_0=(2\mu_0/\pi\eta)/[1+(\mu_0/\eta)^2]$,
 which we see extends to $\Omega\lesssim 2\mu$ and, as noted before, the main rise towards
the universal background value is set by $2\mu$ rather than by $\eta$.
The few intermediate curves show the evolution from one regime to the other.

\section{Infrared Conductivity}

\begin{figure}[ht]
\begin{picture}(250,200)
\leavevmode\centering\includegraphics{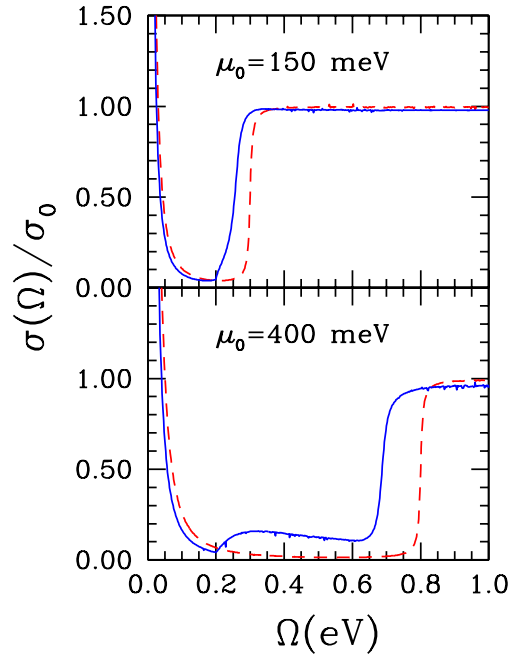}
\end{picture}
\vskip 40pt
\caption{(Color online) AC conductivity $\sigma(\Omega)/\sigma_0$
for two values of chemical potential $\mu_0=150$ meV (top frame)
and $\mu_0=400$ meV (bottom frame), one below and one above the
phonon energy of $\omega_E=200$ meV. The dashed red curve was obtained
without electron-phonon renormalization and the solid blue is with. 
}
\label{Fig16}
\end{figure}

So far we have examined the small photon energy range of the AC conductivity
as compared with the value of the 
 phonon energy $\omega_E=200$ meV in our model. In
Fig.~\ref{Fig16}, we show results to $\Omega=1$ eV for $\mu_0=150$ meV
(top frame) and $\mu_0=400$ meV (bottom frame). The dashed red curve,
which is included for comparison,  give results without
the electron-phonon interaction (bare band case) but  with a 
phenomenological impurity transport 
scattering rate of $2\eta$ with $\eta=2.5$ meV 
(constant). The solid blue curve has phonons. The first thing to note
when comparing these curves is that the main rise in
$\sigma(\Omega)/\sigma_0$,
indicating the increase in interband transitions, occurs at twice the
value of the dressed chemical potential in the solid curve while
it occurs at $2\mu_0$ for the bare bands. This translates into a
considerable
shift downward of this prominent threshold for absorption. This is
easily understood with the help of Fig.~\ref{Fig10} where bare
band
and renormalized dispersion curves are shown. Absorption of light for
the interband case proceeds through vertical transitions from 
an occupied state in the lower
Dirac cone to an unoccupied state in the upper Dirac cone,
since
the photon transfers no momentum to the electronic system.
 This is
shown
as the arrows which applies to the lowest energy transition
allowed by the Pauli exclusion principle, which blocks transitions below 
this energy. 
For the bare clean bands this energy is just $2\mu_0$ (large black arrow),
twice the bare chemical potential. But in the interacting band the
first
transition occurs rather at $2\mu$ as shown on the left hand side of
the figure by a blue vertical arrow which is clearly shorter in length
than is the black arrow. In reality, of course, there are finite
lifetime
effects which broaden the initial and final states (except for the one
right at the Fermi energy) and this is reflected in a rounding of the
absorption edge. In our numerical work for bare bands, we have
included a small residual scattering rate on our states so that even
for the red dashed curve a small rounding of the interband edge is
seen
in the figure. Note that this shift is also present in the previous
Figs.~\ref{Fig14} and \ref{Fig15}.

\begin{figure}[ht]
\begin{picture}(250,200)
\leavevmode\centering\includegraphics{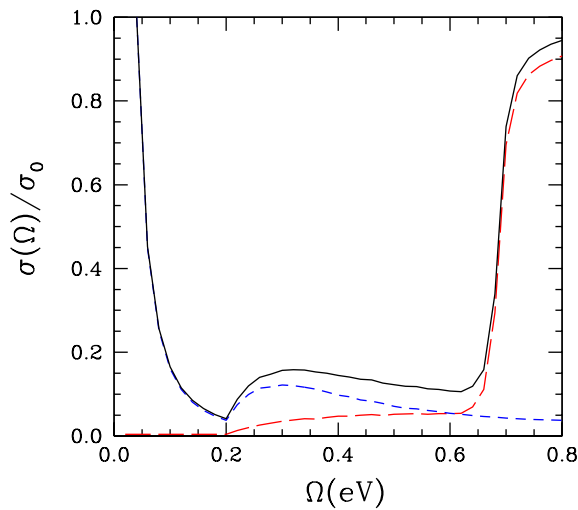}
\end{picture}
\vskip 20pt
\caption{(Color online) AC conductivity $\sigma(\Omega)/\sigma_0$ showing 
separately intraband (short-dashed blue curve) and interband (long-dashed
red curve). The solid black curve is the sum of the two. The chemical
potential $\mu_0=400$ meV. 
}
\label{Fig17}
\end{figure}

\begin{figure}[ht]
\begin{picture}(250,200)
\leavevmode\centering\includegraphics{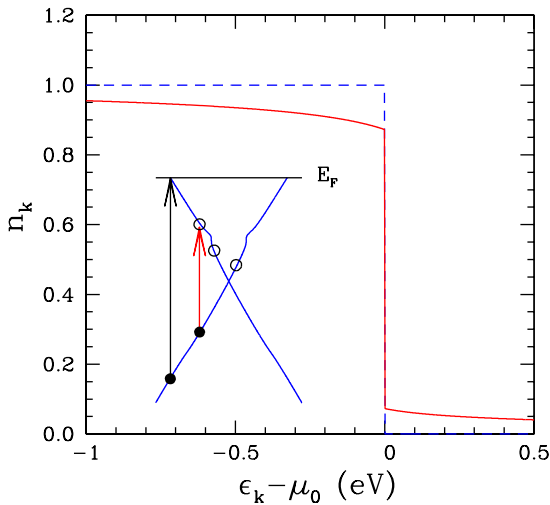}
\end{picture}
\vskip 20pt
\caption{(Color online) Probability of occupation of state $\epsilon_k$
for two cases: bare band (dashed blue curve) and with electron-phonon
interaction included (solid red curve). A very small residual scattering
of $\eta=0.001$ eV is also present in both cases. The inset is a schematic
which illustrates the renormalized energy bands 
filled to the Fermi level $E_F$ with finite probability for
some holes to exist below the Fermi level.
Interband transitions are now possible for energies below $2\mu$.
}
\label{Fig18}
\end{figure}

Another feature to be noted when considering Fig.~\ref{Fig16},
is the contrast between top and bottom frame for which $\mu_0$ goes
from a value of 150 meV, which is less than $\omega_E=200$ meV,
to a value of $\mu_0=400$ meV which is larger. The large value of
$\mu_0$
allows the Holstein phonon-assisted sideband to be prominently revealed.
It starts at $\Omega=200$ meV in the blue curve, while for the bare
band
it is not present. What is important to realize is that both
interband
and intraband processes contribute to this absorption. The intraband
contribution is easily understood from conventional metal theory and
arises when a photon not only induces an electronic transition, i.e.
creates an excited hole-particle pair, but also creates a phonon
in the final state.
But there is also an interband contribution to the Holstein sideband
as can be seen in Fig.~\ref{Fig17}, where the total conductivity
(solid black curve) is decomposed into an intraband (short-dashed blue
curve) and an interband (long-dashed red curve) contribution. To
understand
the behavior of the interband curve we refer to
Fig.~\ref{Fig18},
where we show results for the probability of occupation of a state
${\bf k}$, $n_{\bf k}$,  which can easily be calculated from the spectral density
$A({\bf k},\omega)$
of Eq.~(\ref{eq:spectral}). At finite temperature
\begin{equation}
n_{\bf k}=\int^\infty_{-\infty}f(\omega)A({\bf k},\omega)d\omega .
\end{equation}
If, for the moment, we ignore damping effects, the electron spectral
density $A({\bf k},\omega)$ becomes a delta function and the probability
of occupation of a state $(\epsilon_k-\mu_0)$ in the bare band becomes
$n_{\bf k}=f(\hbar v_0(k-k_F))$  for the
$\epsilon_{\bf k}=+\hbar v_0 k$ branch and
$n_{\bf k}=f(-\hbar v_0(k+k_F))$  for the
$\epsilon_{\bf k}=-\hbar v_0 k$ branch,
which is just the thermal occupation
function. When the EPI is included, an interesting analytical result can
still be obtained if we limit ourselves to the region of small energies
away from the Fermi surface. In that case, we can approximate ${\rm Re}\Sigma(\omega)$
by Eq.~(\ref{eq:15}) and obtain
\begin{equation}
n_{\bf k}\simeq\frac{1}{1+\lambda^{\rm eff}}f(\hbar v_0^*(k-k_F)),
\label{eq:nkapprox}
\end{equation}
where $v_0^*$ (defined below Eq.~(\ref{eq:16}))
is the renormalized effective velocity of light, $k_F$
is the bare band Fermi momentum and the equation is valid only for 
$(k-k_F)$ small. We can immediately see from (\ref{eq:nkapprox})
that the jump in occupation at the Fermi surface in the interacting
system at $T=0$ has been reduced below one by the factor $1/(1+\lambda^{\rm eff})$.
Our numerical results with and without phonons are shown in
Fig.~\ref{Fig18}.
The dashed blue curve is close to unity except very near the Fermi surface
at $k=k_F$ in our notation. We have used a residual scattering 
$\eta=0.001$ meV.
When phonons are included, we obtain the solid red curve
which also shows a sharp drop at the Fermi surface, but the jump 
is now reduced below one by a factor,
$1/(1+\lambda^{\rm eff})$, as we saw in
Eq.~(\ref{eq:nkapprox}). 
In addition, this curve has very
significant
tails beyond the Fermi level and at negative energies below this point,
the probability of occupation remains reduced from the
value
of 1 on the scale of eV
reflecting the fact that phonon renormalizations persist to high energies
in graphene as we saw in Fig.~\ref{Fig1}. 
Thus since $n_{\bf k}$ is always less than 1, Pauli blocking is
lifted on interband transitions. The initial state in the occupied
lower Dirac cone of Fig.~\ref{Fig10} remains occupied  with finite
probability
but at the same time a final state below the Fermi level can now
accommodate
a photo-excited  electron since this state is not occupied
with probability one. This is illustrated by the inset schematic of
Fig.~\ref{Fig18}, where the large black arrow represents the
case where the particle is promoted to above the
Fermi level, which would be a transition with minimum energy $2\mu$.
However, the finite probability for holes to exist below the Fermi level
gives rise to transitions such as shown by the short red arrow, and 
interband absorption can now proceed for energies which are less that
$2\mu$.
This partial lifting of Pauli blocking is
responsible
for the long tails extending to $\Omega=0$ seen in the lower right
hand
frame of Fig.~\ref{Fig14} which allow the interband transitions
to contribute to the DC conductivity. They also allow for additional
phonon-assisted interband transitions to start at $\Omega>\omega_E$
as seen in the long-dashed red curve of Fig.~\ref{Fig17}.
An interesting point to be aware of is that for large values of the
chemical potential, a regime in which a Drude and its boson-assisted
side bands is clearly revealed, the sidebands will not have the same
relationship to the Drude as they would in a simple metal. We expect
an additional contribution in this energy range from interband
transitions, as shown in Fig.~\ref{Fig17}.
We saw in Eq.~(\ref{eq:nkapprox}) that the jump at $k=k_F$ in the
occupation probability is renormalized by the EPI by a factor
of $1/(1+\lambda^{\rm eff})$. We also saw in Fig.~\ref{Fig18} that these
renormalizations persist to high energy away from the Fermi surface
and the occupation factor $n_{\bf k}$ is significantly reduced below
one allowing for the relaxation of Pauli blocking on interband transitions.
On the other hand, these renormalizations drop out of the value of the 
interband conductivity in Eq.~(\ref{eq:condintra}) valid for small photon
energy and Fig.~\ref{Fig15} shows this to be true for a considerable range of
energy above $\Omega=0$. This is a clear illustration that the EPI can
renormalize various quantities in quite different ways. 

\begin{figure}[ht]
\begin{picture}(250,200)
\leavevmode\centering\includegraphics{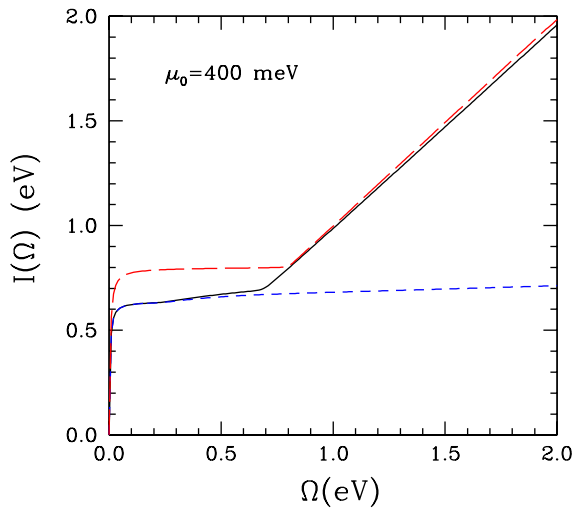}
\end{picture}
\vskip 20pt
\caption{(Color online) The partial optical sum $I(\Omega)$ defined in
Eq.~(\ref{eq:opsum}) vs $\Omega$ (variable upper limit on the integral).
The red long-dashed curve  is for the bare band case with $\eta=2.5$ meV
and is for comparison. The solid black curve  has phonon renormalizations.
The short-dashed blue curve includes only the intraband contribution in
the presence of phonon renormalizations.
}
\label{Fig19}
\end{figure}

Another interesting aspect of the redistribution of optical spectral
weight that is brought about by the electron-phonon interaction is
illustrated in Fig.~\ref{Fig19} where we show the results for the 
partial optical sum denoted by $I(\Omega)$ and defined as 
\begin{equation}
I(\Omega)=\int_0^\Omega\frac{\sigma(\Omega')}{\sigma_0}d\Omega' ,
\label{eq:opsum}
\end{equation}
with variable upper limit $\Omega$ on the integral.
The red solid curve is the bare band case but with a small residual
quasiparticle 
scattering rate $\eta=2.5$ meV included in the numerical evaluation of
Eq.~(\ref{eq:cond}). In this case, the entire spectral weight $2\mu_0$ lost through
Pauli blocking of the interband transition is found in the
Drude. $I(\Omega)$
rises sharply to a height of 0.8 eV on an energy scale given by $2\eta$,
then
remains constant until $\Omega=0.8$ eV $=2\mu_0$  where it starts
rising again in a linear fashion. The solid black curve which includes
phonons is more interesting and is to be compared with the red curve.
It too rises out of zero very rapidly on the scale of $\eta$ but does
not
go all the way up to a height of $2\mu_0$ rather it shows a saturation plateau
at $2\mu_0/(1+\lambda^{\rm eff})=2\mu$ in the region below $\omega_E=200$
meV.
At $\omega_E$, intra and interband Holstein processes start to kick in
and $I(\Omega)$ begins to rise slowly until $\Omega\sim 2\mu$ is
reached
where a considerable change in slope occurs because the major
contribution of interband transitions starts coming in (as can
be seen by comparing the blue short-dashed intraband curve with the
solid black curve for the total amount).
By $\omega=2\mu_0$, the black and red curves have effectively merged
and the optical spectral weight redistribution brought about by the EPI
is in balance at this point.
At larger energies $\Omega\sim 2$ eV, the solid black curve which
is based on the renormalized conductivity remains below the solid red curve.
This is because the electron-phonon interaction has reduced the value
of the universal background slightly below $\sigma_0=\pi e^2/2h$ (also
seen in Fig.~\ref{Fig16} at large $\Omega$, where the solid blue curve
is
slightly below the red dashed curve).
This is
understandable. We saw in Fig.~\ref{Fig5} that the electron-phonon
interaction
has a profound effect on the band structure in the energy region
around
the band edge. In this region, the DOS is considerably depleted below
its noninteracting value and to conserve states tails appear beyond
the bare cut off $W_C$. Thus in optical experiments, spectral weight is
removed below the bare optical cut off which is transferred to higher
energies. The slight 
reduction below one of the universal background is generic and we would
expect it to be a feature of interactions in general.

\section{Summary and Conclusions}

It is well known\cite{Grimvall,Prange}
that in wide band metals for which the band density of states
(DOS) is essentially constant on the scale of a few times a
phonon energy, the electron-phonon interaction drops out of the 
renormalized DOS. Because of the Dirac nature of electron dynamics
in graphene the bare band DOS is linear in energy rather than constant
and consequently an image of the phonons is retained in its 
dressed DOS and a first derivative of $N(\omega)$
provides an ideal baseline to study boson structures.
For a bare Dirac band this derivative gives minus or plus one with the
change in sign occurring at $-\mu_0$, i.e. at minus the bare chemical
potential which also coincides with the Dirac point.
Several modifications arise when the electron-phonon interaction is
included. First, the Dirac point position is shifted and this leads 
to a change in the energy of the jump from minus to plus one in the first
derivative. This jump remains sharp provided the chemical potential
is in magnitude smaller than the phonon energy. But when
the electron-phonon lifetime is finite at the Dirac point, the jump 
becomes smeared. 
 In addition
there are structures at $\pm\omega_E$ for an Einstein spectrum.
When a distributed phonon spectrum is used, the
structure  in $dN(\omega)/d\omega$ reflects not only its
magnitude but also it shape. With increasing doping or more precisely 
magnitude of chemical potential, the phonon structures are enhanced
because the Fermi level falls in a region of higher density of states
and consequently of larger strength of the effective electron-phonon
interaction. 
A point of note is that at the Dirac point, the explicit logarithmic
phonon structures of the finite doping case disappear, although the DOS
still retains its $(1+\lambda^{\rm eff})$ mass renormalization. Small
features
remain in the derivative and there is some evidence from STM\cite{li}
for both of these effects in this limit.
The method should be applicable to other
scattering mechanisms such as electron-hole and plasma excitations.

Another relevant observation is that in the DOS, we find that for
$\mu_0>\omega_E$, the interactions lift the Dirac point from a value of 
 zero in the bare case to a finite value, and that the DOS rises quadratically
rather than linearly out of this point. This is due to a finite scattering
rate at the Dirac point.
Recent STM experiments\cite{zhang} appear to see
this feature of lifting and accompanying quadratic behavior. Further
investigation of this point would be warranted.

We have considered separately the effect of the electron-phonon 
interaction (EPI) on the intra and interband contributions to the
optical conductivity. We find that the DC conductivity remains unrenormalized
with  equal contributions from both processes when the chemical 
potential is zero. With increasing doping through application of a gate voltage
in a field effect device or by seeding the graphene surface with
potassium or other atoms, the interband contribution decreases
and becomes negligible when the transport scattering rate $\eta$ becomes
much smaller than the chemical potential. 
The conditions for the observation of an universal DC limit is $\mu\ll\eta$,
as well as, $T\ll\eta$. For $T>\eta$, there is a rapid increase in DC conductivity
in contrast to ordinary band metals for which it is independent of temperature.
For zero doping, the scale on which the universal DC value $4e^2/\pi h$
rises with increasing photon energy to its universal background value of
$\pi e^2/2 h$  is set by the transport scattering rate $\eta$.
By contrast, but in complete
parallel to what is known to apply in simple band metals, 
as the chemical potential is increased, a clear AC intraband Drude conductivity
is revealed which is renormalized in two ways.
 Through analytic
techniques, we show that the transport scattering rate
$1/\tau$ is to be replaced by
$1/[\tau(1+\lambda^{\rm eff})]$ and the plasma frequency
squared
reduced to $\Omega_p^2/(1+\lambda^{\rm eff})$. Here, $\lambda^{\rm
eff}$
is the electron effective mass renormalization at the Fermi energy
which varies with doping.
In addition, optical spectral weight is transferred to higher energy
in the form of a Holstein phonon-assisted absorption sideband. Also,
interband transitions, not part of the theory of simple metals,
provide additional absorption.
At small frequencies below twice the value of the chemical
potential $(\mu)$ there is a small but finite nearly frequency
independent interband absorption which shows an additional Holstein
side band for the case $2\mu>\omega_E$ (the phonon energy), with an
additional rapid increase towards its universal background
value $\sigma_0=\pi e^2/2h$ at $\Omega\gtrsim 2\mu$. The near
constant interband absorption below $2\mu$ can be traced to a lifting of Pauli
blocking brought about by the EPI. The interactions decrease the probability
of occupation of a plane wave state below the Fermi energy to a value
less than one so that such states can still be used as final states for
interband transitions although these are greatly reduced in optical
spectral weight. In pure graphene, the increase in conductivity at 
$\Omega=2\mu_0$ represents a vertical jump from 0 to $\sigma_0$. When the
EPI is included not only does the chemical potential shift to a smaller
value but the jump at $2\mu$ is also smeared due to
finite lifetime effects. While at small frequencies in the Drude region
the conductivity can exceed its background value $\sigma_0$, for energies
greater than $2\mu$ it is always slightly {\it smaller} than $\sigma_0$. Consideration
of the partial optical sum rule shows that missing optical spectral
weight
at high energy
is transferred to even higher energies as compared to 
the bare band case because the EPI
provides states beyond the bare band cutoff on the scale of the phonon energy.
It also shows that at small energies the readjustment of optical
spectral weight between bare and interacting case is in near balance
at $\Omega\simeq 2\mu_0$.

\begin{acknowledgments}
This work has been supported by NSERC of Canada (E.J.N. and J.P.C.)
and by the Canadian Institute for Advanced Research (CIFAR) (J.P.C.).
S.G.S. was supported by the Program of Fundamental Research of Physics
and Astronomy Division of the National Academy of Sciences of the Ukraine
and thanks V.P. Gusynin for many stimulating discussions.
\end{acknowledgments}

\end{document}